\documentclass[aps,floats,twocolumn,epsf,prl,showpacs,superscriptaddress]{revtex4-1}
\usepackage{graphicx}
\usepackage{epstopdf}
\usepackage{amsmath}
\usepackage{amssymb}
\usepackage[colorlinks=true,urlcolor=blue,citecolor=blue,linkcolor=blue,breaklinks=true]{hyperref}
\usepackage{color}
\usepackage[normalem]{ulem}
\usepackage[dvipsnames]{xcolor}

\usepackage{ulem}

\begin{document}

\title{Luttinger surface and exchange splitting induced by ferromagnetic fluctuations}

\author{Motoharu Kitatani}
\affiliation{Department of Material Science, University of Hyogo, Ako, Hyogo 678-1297, Japan}

\author{Yusuke Nomura}
\affiliation{Institute for Materials Research (IMR), Tohoku University, Katahira, Aoba-ku, Sendai 980-8577, Japan}
\affiliation{Advanced Institute for Materials Research (WPI-AIMR), Tohoku University, Katahira, Aoba-ku, Sendai 980-8577, Japan}

\author{Shiro Sakai}
\affiliation{Physics Division, Sophia University, Chiyoda-ku, Tokyo 102-8554, Japan}
\affiliation{RIKEN Center for Emergent Matter Sciences (CEMS), Wako, Saitama, 351-0198, Japan}

\author{Ryotaro Arita}
\affiliation{RIKEN Center for Emergent Matter Sciences (CEMS), Wako, Saitama, 351-0198, Japan}
\affiliation{Department of Physics, The University of Tokyo, Hongo, Tokyo, 113-8656, Japan}

\date{\today}

\begin{abstract}
Ferromagnetism in the single-orbital Hubbard model, which contains only local Coulomb repulsion and no explicit ferromagnetic exchange interactions, has been extensively studied. However, how the associated fluctuations influence the electronic properties near the transition remains a fundamental issue. Here, by applying the dynamical vertex approximation (D$\Gamma$A) to single-orbital systems with a partially flat band dispersion, we demonstrate that finite-correlation-length ferromagnetic fluctuations generate an emergent Luttinger surface and drive a Fermi surface expansion reminiscent of exchange splitting, even without magnetic order. We further derive an analytical expression that reproduces these effects, clarifying the microscopic origin for fluctuation-driven exchange splitting in correlated electron systems.
\end{abstract}

\maketitle


{\it Introduction.}---
In real itinerant ferromagnets such as transition metals and their compounds, theoretical descriptions are most often based on multi-orbital models, where Hund's coupling plays a decisive role in forming large local moments and stabilizing ferromagnetic order \cite{Momoi1998,Held1998,Lichtenstein2001,Sakamoto2002,Sakai2007}. By contrast, a single-orbital Hubbard model—the most stripped-down representation of correlated electrons—lacks explicit ferromagnetic exchange, and its tendency toward ferromagnetism has long been investigated \cite{Kanamori1963,Nagaoka1966,Mielke1991,Tasaki1992}. Beyond the stability itself, an equally important issue is how the associated fluctuations influence the electronic properties in such simplified settings.

Several rigorous results highlight specific scenarios: Nagaoka ferromagnetism proves complete spin polarization at infinite $U$ with a single doped hole \cite{Nagaoka1966}, and flat-band ferromagnetism demonstrates that perfect band flatness engineered by destructive interference in hopping can stabilize a ferromagnetic ground state even at finite $U$ \cite{Mielke1991,Tasaki1992}. Subsequent studies have shown that ferromagnetism in flat-band systems survives even when the band is not perfectly flat \cite{Tasaki1995,Hlubina1997,Hlubina1999,Irkhin2001,Honerkamp2001a,Honerkamp2001b}. These states usually occur at low filling, where their nature may not be the same as multi-orbital ferromagnetism. This makes it particularly important to investigate the intrinsic properties of such single-orbital models. Indeed, motivated by this perspective, spin-wave excitations have been examined \cite{Kusakabe1994,Kusakabe1994b}.

In strongly-correlated systems, fluctuations in the vicinity of a phase transition often cause nontrivial spectral reconstructions. Experiments on Fe thin films \cite{Pickel2010}, SrRuO$_3$ thin films \cite{Jeong2013,Hahn2021}, and cobalt \cite{Eich2017} have reported exchange-splitting-like features and shadow bands persisting above the Curie temperature. On the theoretical side, the two-dimensional Hubbard model with next-nearest-neighbor hopping has revealed strong ferromagnetic instabilities and spectral anomalies, including band splitting and non-Fermi-liquid behavior near the transition \cite{Monthoux2003,Hankevych2003,Katanin2005,Katanin2005b,Katanin2008,Sayyad2020,Sayyad2023}. In multi-orbital systems, cluster extensions \cite{Lichtenstein2000,Kotliar2001,Maier2005} of dynamical mean-field theory (DMFT) \cite{Metzner1989,Georges1992,Georges1996} have shown that the Fermi surface volume expands already above $T_{\rm c}~$\cite{Nomura2015,Nomura2022}. Taken together, these results suggest that ferromagnetic correlations have a substantial impact on the electronic spectrum beyond the Stoner picture, yet their microscopic nature and influence on the spectrum remain elusive. Just as antiferromagnetic fluctuations play an important role in the pseudogap phase of cuprates \cite{Keimer2015,Vilk1997,Borejsza2004,Schaefer2021,Schmalian1999,Schaefer2015,Krien2022}, ferromagnetic fluctuations are expected to play an equally significant role in real materials, which motivates further investigation using recent advances in many-body methods.

In this Letter, we investigate this problem using the dynamical vertex approximation (D$\Gamma$A) \cite{Toschi2007,Katanin2009,Kusunose2006}, a state-of-the-art nonperturbative method extending DMFT to treat both local and long-range fluctuations. Applying D$\Gamma$A to the two-dimensional single-orbital Hubbard model with partially flat bands, we demonstrate that above the Curie temperature, exchange splitting emerges, a Luttinger surface forms, and the Fermi surface volume expands as if long-range ferromagnetic order were present. Strikingly, these effects arise solely from finite-correlation-length fluctuations, without any explicit ferromagnetic exchange interaction or multi-orbital physics. We further derive an analytical expression that quantitatively reproduces the numerical results and clarifies the role of band curvature in amplifying these effects. Our findings reveal a microscopic origin for exchange splitting in correlated metals, provide a theoretical basis for interpreting anomalous spectral reconstructions observed in experiments, and point toward unexplored routes to exotic states in flat-band systems.

\noindent
{\sl Model and Methods}---We study the two-dimensional Hubbard model on the square lattice as
\begin{align}
    {\cal H}= \sum_{\mathbf{k},\sigma}
    \epsilon_{\mathbf{k}} c^{\dag}_{\mathbf{k},\sigma} c_{\mathbf{k},\sigma}^{\phantom{\dag}}
    +U \sum_{i} n_{i,\uparrow}n_{i,\downarrow},
\end{align}
where $c^{\dag}_{\mathbf{k},\sigma}(c_{\mathbf{k},\sigma}^{\phantom{\dag}})$ is the creation (annihilation) operator of the electron with spin $\sigma=\uparrow,\downarrow$, $n_{i,\sigma}\equiv c^{\dag}_{i,\sigma}c_{i,\sigma}$ the number operator at site $i$, 
\begin{align}
    \epsilon_{\mathbf{k}}=
    -2t[{\rm cos}(k_x)+{\rm cos}(k_y)]
    -4t^{\prime}{\rm cos}(k_x){\rm cos}(k_y) 
\end{align}
the energy-momentum dispersion, $U$ the onsite Coulomb repulsion, and $t,\ t^{\prime}$ are the nearest and second nearest neighbor hoppings, respectively. 
We focus on the system with the large next nearest neighbor hopping ($t'/t=-0.4 \sim -0.5$) so that the system has a high density of states on the low-energy side. We focus on the low-density regime where the Fermi level is near the van Hove point. This system has been discussed to have large ferromagnetic instability \cite{Hlubina1997,Hlubina1999,Irkhin2001,Honerkamp2001a,Honerkamp2001b,Hankevych2003,Katanin2005}.

To analyze this model, we use
D$\Gamma$A~\cite{Toschi2007,Katanin2009,Kusunose2006}, which is a diagrammatic extension of DMFT \cite{Metzner1989,Georges1992,Georges1996}; for details, see Refs. \cite{Rohringer2018,DGAdetails}. 
This method can simultaneously treat local and long-range fluctuations. D$\Gamma$A has captured the pseudogap and superconductivity in cuprate and nickelate superconductors with strong antiferromagnetic fluctuations \cite{Kitatani2019,Kitatani2020,Kitatani2023}. Also, this method has clarified the quantum critical phenomena where the fluctuation is extremely strong \cite{Rohringer2011,Schaefer2017,Schaefer2019,Kitatani2025}, revealed particle-hole asymmetric lifetimes induced by critical fluctuations \cite{Pickem2022}, and was benchmarked by the numerically exact diagrammatic Monte Carlo \cite{Schaefer2021}. For comparison, we also present fluctuation exchange (FLEX) \cite{Bickers1989} and cluster DMFT (cDMFT) \cite{Lichtenstein2000,Kotliar2001,Maier2005} results in the Supplemental Material \cite{SM}.

\begin{figure}[tbp]
        \centering
        \includegraphics[width=\linewidth,angle=0]{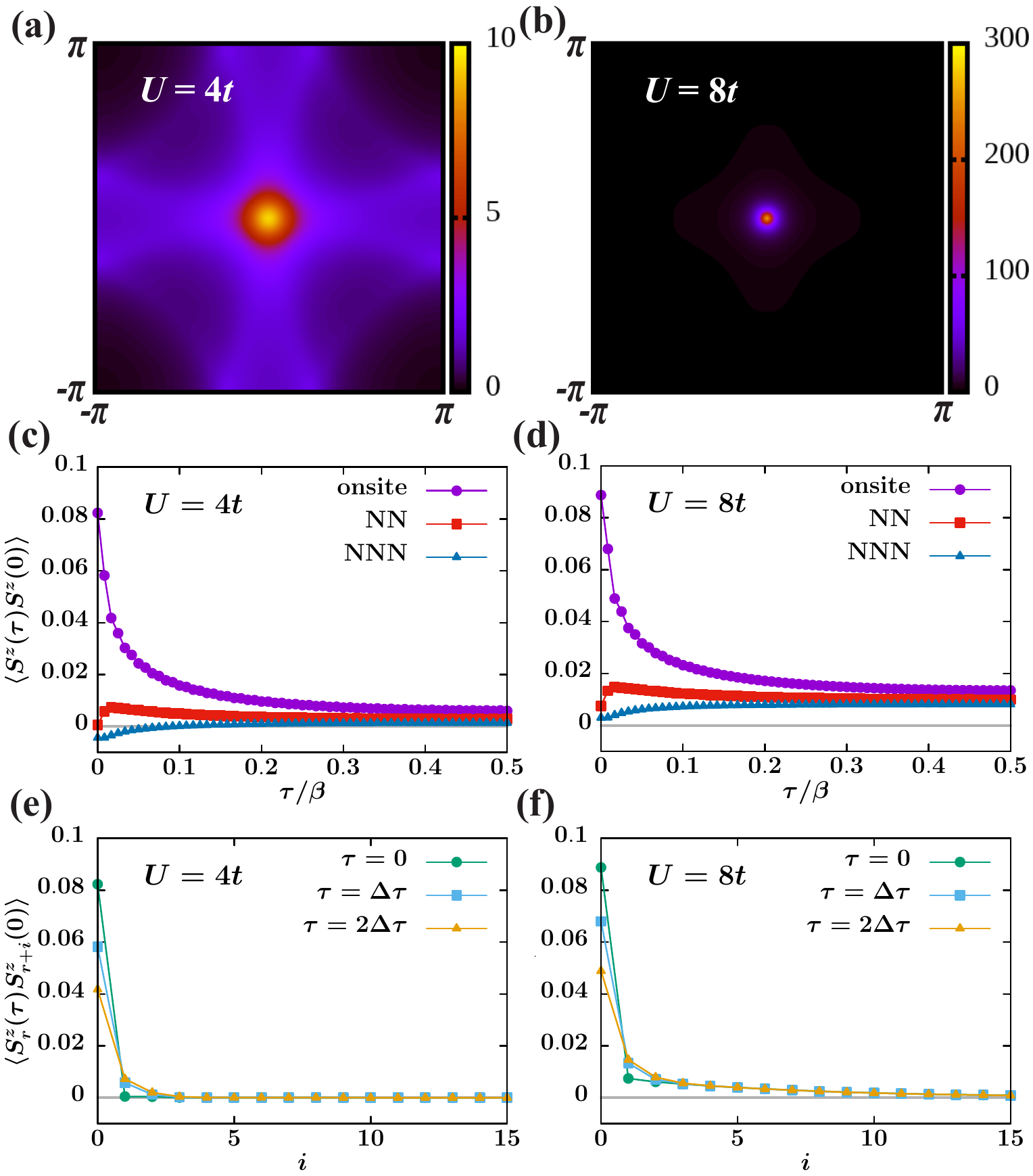}
        \caption{Momentum dependence of the spin susceptibility $\chi_{\rm s}(\mathbf{k},\omega=0)$ for (a) $U=4t$ and (b) $U=8t$ and (c,d) the imaginary time $\tau$ and (e,f) the spatial dependence of the spin-spin correlation function at $t^{\prime}=-0.45t,\ T=0.02t,\ n=0.40$.}
        \label{fig:spinfluc}
\end{figure}

{\it Spin fluctuations.}---Let us first show how the spin susceptibility develops in this model. Figures \ref{fig:spinfluc}(a,b) represent the momentum dependence of the spin susceptibility $\chi_{\rm s}(\mathbf{k})=2\int_0^{\beta}d\tau \langle S^{z}_{\mathbf{k}}(\tau)S^{z}_{-\mathbf{k}}(0) \rangle$ at $t^{\prime}=-0.45t, T=0.02t$ for (a) $U=4t$ and (b) $U=8t$. For both cases, the spin susceptibility peaks at around $\mathbf{k}=(0,0)$, manifesting ferromagnetic instability. In Fig.~\ref{fig:spinfluc}(a), we can see another ``+"-shaped structure of the susceptibility, which reflects the ``partially flat-band" \cite{Huang2019,Sayyad2020} nature of the model (i.e., possible particle-hole excitations for wide momentum regions). For a stronger coupling regime, we observe only the strong ferromagnetic fluctuations around $\mathbf{k}=(0,0)$ and other structures are negligible.
In Fig.~\ref{fig:spinfluc}(c,d), we show imaginary time ($\tau$) dependence of the spin-spin correlation function for onsite, nearest-neighbor (NN) site, and next-nearest neighbor (NNN) site at the same parameter. Similar to the previous cDMFT study of strong Hund's coupling multiorbital systems \cite{Nomura2022}, we obtain the frozen moment behavior in the strong coupling regime. We also show spatial ($i$) dependence of the spin-spin correlation function at $\tau=0,\  \Delta\tau,\  2\Delta\tau$ ($\Delta \tau \equiv \beta/120$, where $\beta$ is the inverse temperature). There is no clear nesting of the Fermi surface, unlike the antiferromagnetic fluctuation around the half filling at $t'/t \simeq 0$. Thus, the spin-spin correlation function decays rapidly compared with the antiferromagnetic fluctuation case \cite{Schaefer2015} (while the tail becomes a bit longer from $U=4t$ to $U=8t$). It is interesting to survey the effect on the self-energy and spectrum from such moderate fluctuations.

\begin{figure}[htbp]
        \centering
        \includegraphics[width=\linewidth,angle=0]{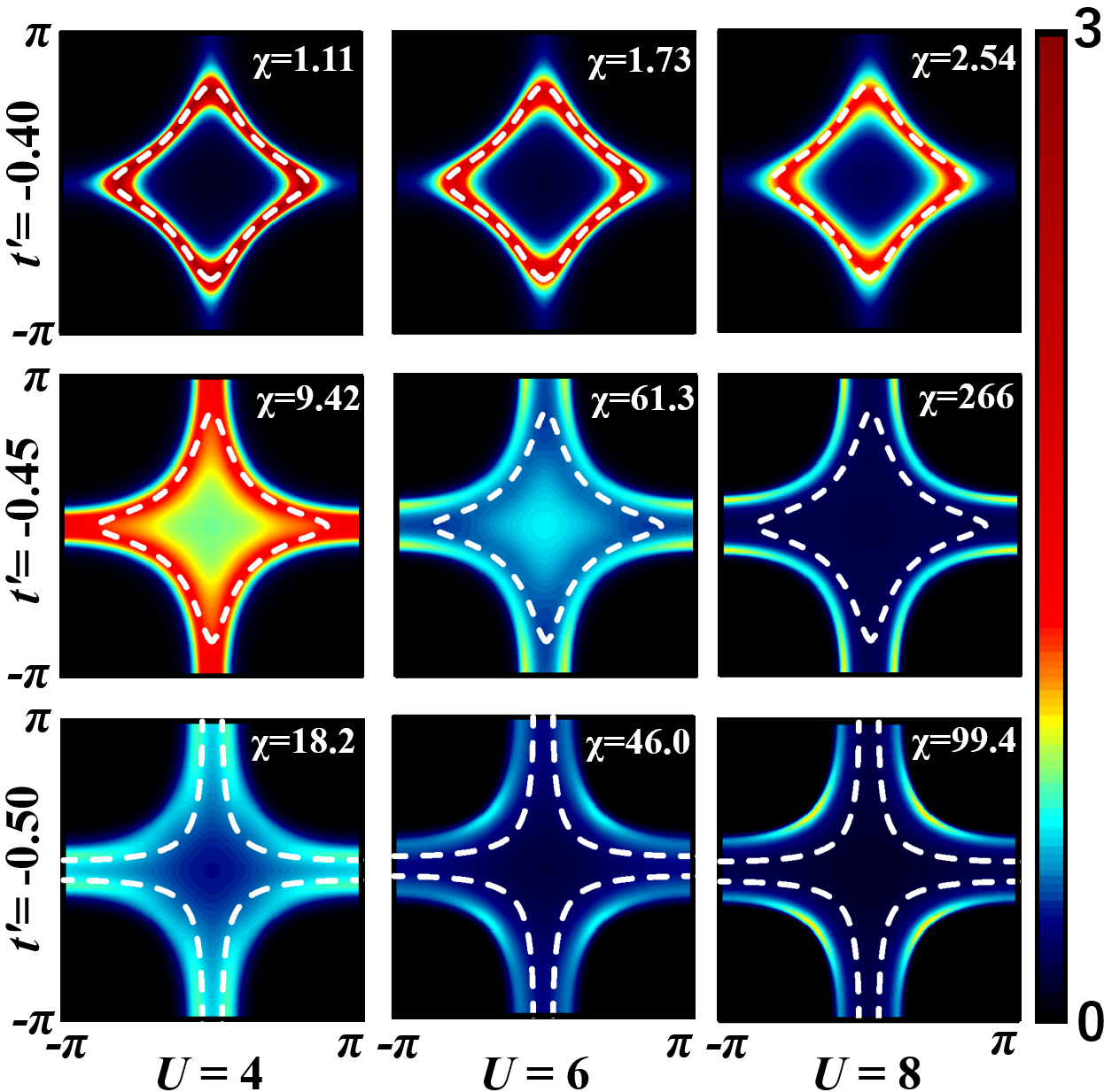}
        \caption{Fermi surfaces (momentum dependence of the imaginary part of the Green's function $-\Im G/\pi$ at the lowest Matsubara frequency) 
        calculated with D$\Gamma$A for $t^{\prime}= -0.40t,\ -0.45t,\ -0.50t$ and $U=4t,\ 6t,\ 8t$ at $T=0.02t, n=0.40$. 
        The value of the ferromagnetic susceptibility $\chi_{\rm s}(\mathbf{k}=0,\omega=0)$ is noted in each figure. Dashed curves correspond the non-interacting Fermi surface for each system.}
        \label{fig:fermisurface}
\end{figure}

{\it Spectrum.}---We next show the imaginary part of the Green's function at the lowest Matsubara frequency ($-\Im G(\mathbf{k},\omega_n=\pi/\beta)/\pi$) for $n=0.40$ in Fig.~\ref{fig:fermisurface} to examine the Fermi surface structure (see the Supplemental Material \cite{SM} for spectra at other fillings). Here, we also show the static spin susceptibility ($\chi_{\rm s}$) and the non-interacting Fermi surface (white dashed curve) for each system. We observe that the spectrum traces the non-interacting band structure when the ferromagnetic fluctuation is weak (e.g., $t^{\prime}=-0.4t$). On the other hand, the Fermi surface expands from the non-interacting shape as the ferromagnetic fluctuation becomes strong, which is the main result of this paper. This effect is most pronounced around $U=8t$ and $t'=-0.45t$ to $-0.50t$, where the ferromagnetic fluctuation is large and the spectral peaks trace the non-interacting Fermi surface with twice the original filling. For the intermediate regime (e.g., $U=6t$ and $t^{\prime}=0.45t$), we also see a signature of the Fermi-surface expansion, but at the same time, there is a finite intensity around $\mathbf{k}=(0,0)$, which we discuss below in detail. Note that we do not obtain such a Fermi surface change in cDMFT (See the Supplemental Material \cite{SM}), which suggests the importance of long-range spatial fluctuations.

Naively, we can regard this as another type of pseudogap, where the spectrum splits (as in a ferromagnetically ordered state) and only the lower branch is occupied, while the system retains SU(2) symmetry. While this is a one-shot calculation for the diagrammatic extension part from the converged DMFT result, pseudogap physics would remain even in a self-consistent calculation \cite{Katanin2005}. Self-consistent cDMFT also observes a similar behavior in multi-orbital systems where a large magnetic moment is stabilized by Hund's coupling \cite{Nomura2022}. 

\begin{figure}[tbp]
        \centering
        \includegraphics[width=\linewidth,angle=0]{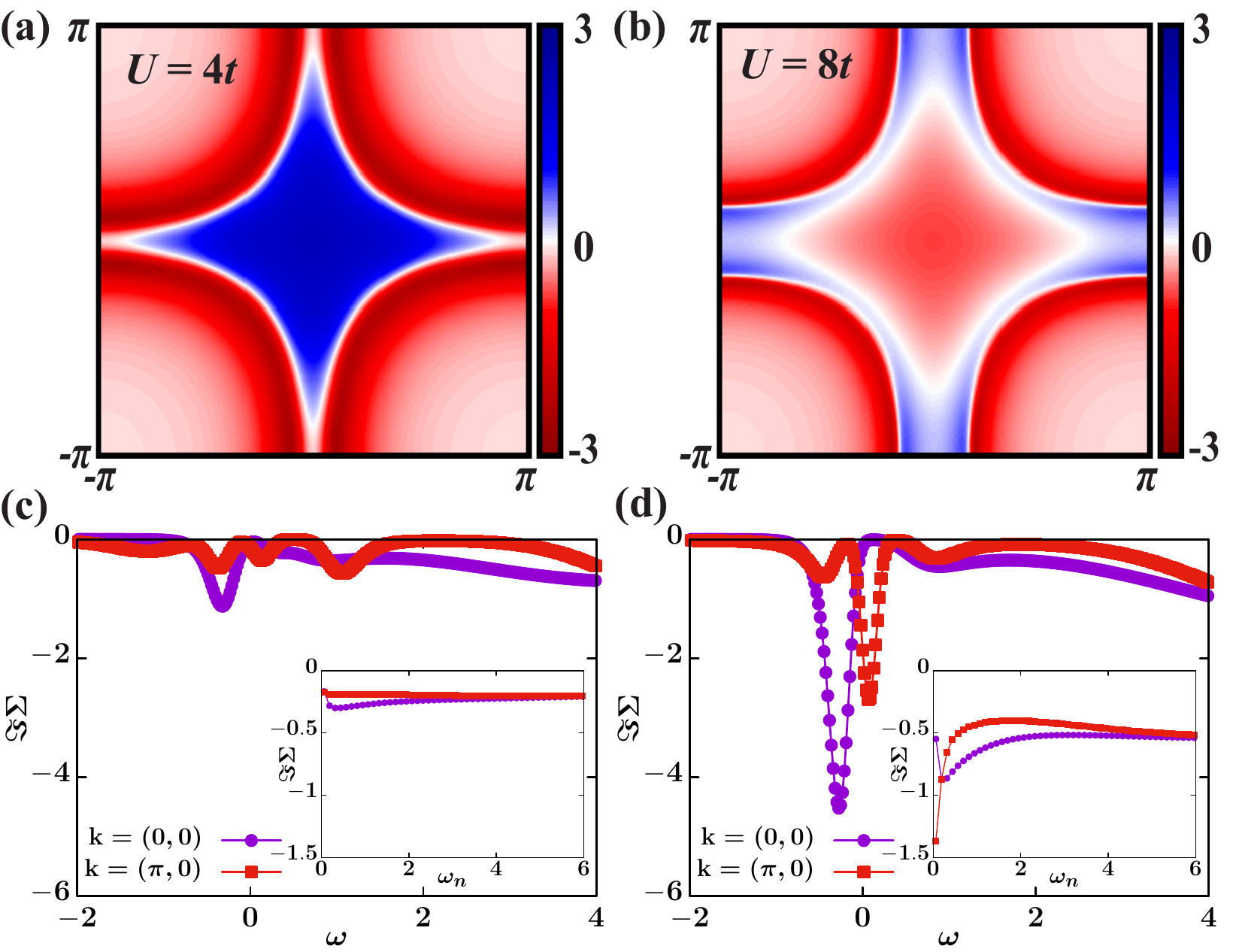}
        \caption{(a,b) Momentum dependence of the real part of the Green's function $\Re G(\mathbf{k},\omega_n=\pi/\beta)$ for $U=4t$ and $8t$ at $t^{\prime}=-0.45t,\  n=0.40$ and $T=0.02t$. (c,d) Real-frequency dependence of the imaginary part of the self-energy for $U=4t$ and $8t$. Insets show the corresponding Matsubara-frequency dependence for the same parameters.
        }
        \label{fig:self-energy}
\end{figure}

In Fig.~\ref{fig:self-energy}(a,b), we show the momentum dependence of the real part of the Green's function $\Re G(\mathbf{k},\omega_n=\pi/\beta)$. For $U=4t$ [Fig.~\ref{fig:self-energy}(a)], the sign of $\Re G$ changes (white curves) only once from $(0,0)$ to $(\pi,\pi)$, reflecting the Fermi surface structure.
In contrast, we obtain the additional $G=0$ line (Luttinger surface) at the Fermi level for $U=8t$ [Fig.~\ref{fig:self-energy}(b)], which should originate from the strong correlation effect in analogy to the Mott physics around half filling \cite{Dzyaloshinskii2003, Stanescu2006,Stanescu2007, Sakai2009,Sakai2023}.
Indeed, we observe a prominent change of the self-energy: For $U=8t$, the imaginary part of the self-energy shows a sharp peak towards $\omega_n \rightarrow 0$ as shown in the inset of Fig.~\ref{fig:self-energy}(d). This is the same trend as the pseudogap opening due to the antiferromagnetic fluctuations around half filling. In contrast, such a self-energy peak doesn't show up for $U=4t$ [inset of Fig.~\ref{fig:self-energy}(c)], where the ferromagnetic fluctuation is weak as mentioned above.
We also show the real frequency dependence of the self-energy in Fig.~\ref{fig:self-energy}(c,d). 
To obtain the real frequency spectrum, we employ the maximum entropy method with the \texttt{ana\_cont} package for the numerical analytic continuation \cite{Kaufmann2023,anacont}. We observe the sharp low-energy peak for $U=8t$, while there are only moderate peaks for $U=4t$. The gap-like feature at $\mathbf{k}=(0,0)$ is rather sharp below the Fermi energy, which is difficult to capture directly from the Matsubara-frequency dependence.

\begin{figure}[tbp]
        \centering
        \includegraphics[width=\linewidth,angle=0]{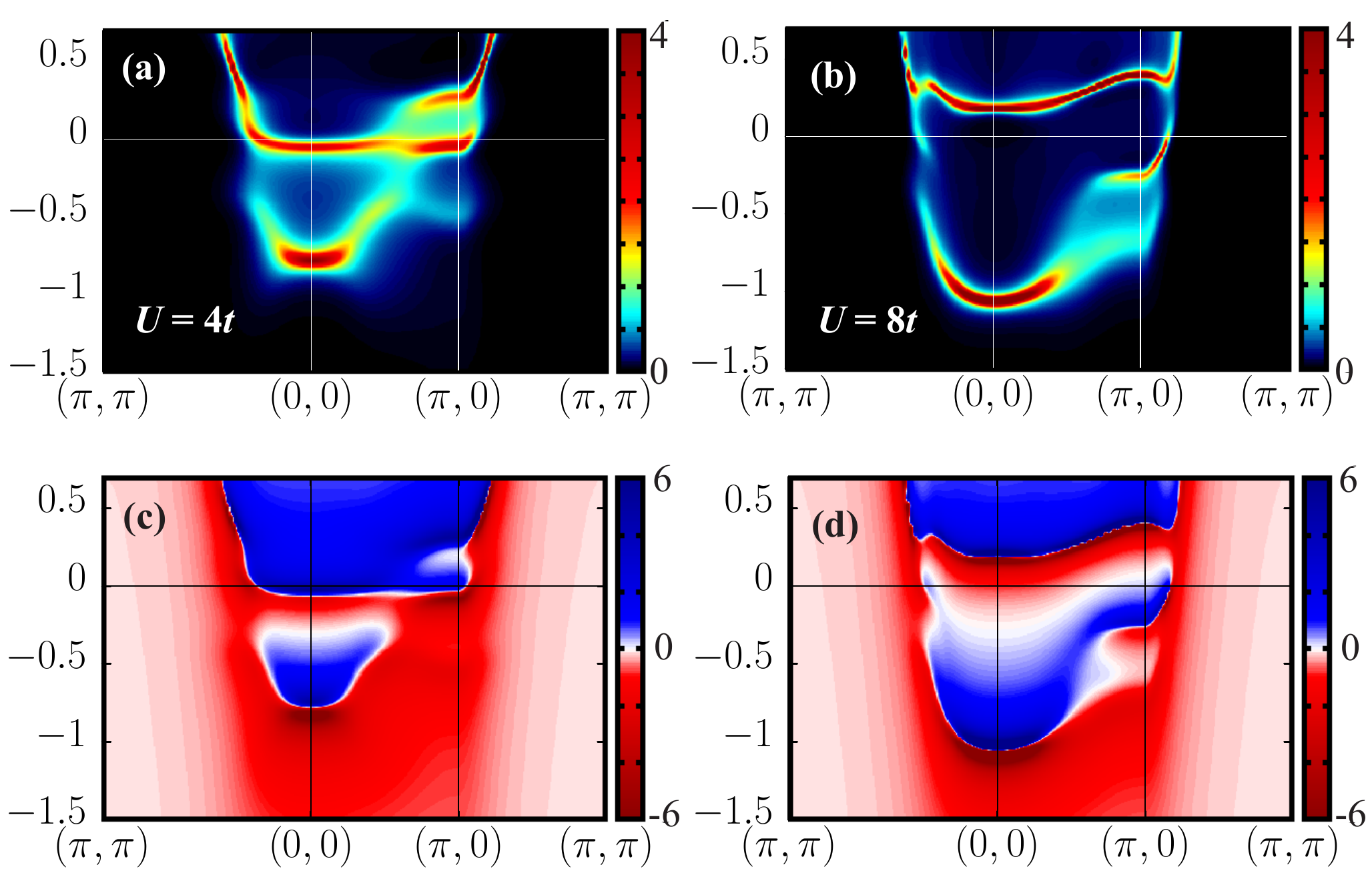}
        \caption{(a,b) D$\Gamma$A spectral weight $A(\mathbf{k},\omega)$ and (c,d) the real part of the Green's function $\Re G(\mathbf{k},\omega)$ for $U=4t$ and $8t$ at $t^{\prime}=-0.45t,\ n=0.40$, and $T=0.02t$.}
        \label{fig:Akw}
\end{figure}

In Fig.~\ref{fig:Akw}, we plot the spectral weight $A(\mathbf{k},\omega)$. For the weaker coupling, $U=4t$, the spectrum splits mainly within the occupied states around $\mathbf{k}=(0,0)$, so the Fermi surface remains almost unchanged. In contrast, for the stronger coupling, $U=8t$, the spectrum splits into occupied and unoccupied parts in the band bottom (partially flat) momentum region.
This fact naturally explains the Fermi-surface expansion observed in Fig.~\ref{fig:fermisurface} because $n_\mathbf{k}$ is smaller than unity in the occupied momentum regime. The spectrum does not fully split into lower and upper branches for unoccupied regions, which we will discuss further in the next section.
We also show the real part of the Green's function in Fig.~\ref{fig:Akw}(c,d). Due to the self-energy peak as discussed above, the $G=0$ line (i.e., Luttinger surface at the Fermi level) appears within the spectral gap. We can see that the $G=0$ appears for only the limited momentum region for $U=4t$, while this clearly appears for a strong coupling regime at $U=8t$. Thus, we attribute the Fermi surface expansion in Fig.~\ref{fig:fermisurface} to the emergence of this Luttinger surface.

At this stage, we can also understand that a certain spectral weight around $\mathbf{k}=(0,0)$ in the intermediate 
fluctuation region
 (e.g., $t^{\prime}=-0.40t$ and $n=0.50$ in Fig.~\ref{fig:fermisurface}) originates from a nearly flat upper band sitting at (or just above) the Fermi level.

{\it Discussion}---
How can we understand the emergence of a Luttinger surface and the accompanying Fermi surface changes?
Under the strong thermal ferromagnetic fluctuation, $W(\mathbf{q},\omega_m) \approx \delta_{m,0}/A(\mathbf{q}^2+\xi^{-2})$, one can evaluate
the one-loop self-energy as $\Sigma(k) = 1/\beta\sum_q W(q)G_0(k+q) \approx G_0(k)/\beta \int_{\mathbf{q}} W(\mathbf{q},\omega_0) = \Delta^2 G_0(k)$ where $k \equiv (\mathbf{k},\omega_n)$ denotes a three-component vector. Here, we assume a divergingly large $\xi$ and neglect the momentum dependence of $G_0(k+q)$ in the summation, and $\Delta^2 \equiv \log(\delta^2\xi^2+1)/(4\pi A\beta)$ where $\delta$ is a cutoff for momentum integration.
Then, we obtain the Green's function as 
\begin{align}
G(k)=\frac{1/2}{i\omega_n-\epsilon_\mathbf{k}-\Delta}
+\frac{1/2}{i\omega_n-\epsilon_\mathbf{k}+\Delta},
\end{align}
which shows a spectral splitting reminiscent of a ferromagnetically ordered state, while a gap appears throughout momentum space in contrast to the numerical results.

To be more precise, the D$\Gamma$A results show that the splitting occurs only in the occupied region. For describing this feature as well, we expand the Green's function around $\mathbf{q}=(0, 0)$ and small momentum modulation effect (basically as in Refs.~\cite{Vilk1997,Schmalian1999,Katanin2005b}, while here we include the second-order contribution, which becomes particularly relevant in the flat-band regime; see below) and  
obtain the self-energy as

\begin{align}
\label{eq:analytical}
\Sigma(k)=&\frac{1}{2\pi^2 A\beta}\frac{\pi G_0(k)}{\sqrt{1+\alpha_k \xi^{-2}}}
\\ \nonumber
&\times
\log{ \left[
\sqrt{\delta^2\xi^2+1}\left(
\frac{\sqrt{1+\alpha_k \xi^{-2}}+1}{\sqrt{1+\alpha_k \xi^{-2}}+\sqrt{1-\alpha_k\delta^2}}
\right)
\right] }, \\ \nonumber
\end{align}
where
\begin{align}
\label{eq:alpha}
\alpha_k \equiv |\nabla \epsilon|^2 G_0(k)^2 + \frac{\Delta \epsilon}{2} G_0(k)
\end{align}
reflects the curvature of the original band structure $\epsilon_\mathbf{k}$ (see the Supplemental Materials \cite{SM} for detailed derivations).
First, we can see that this self-energy goes back to the original form (giving the Green's function in Eq.~(3)) in the limit of $\alpha_k \rightarrow 0$.

\begin{figure}[tbp]
        \centering
        \includegraphics[width=\linewidth,angle=0]{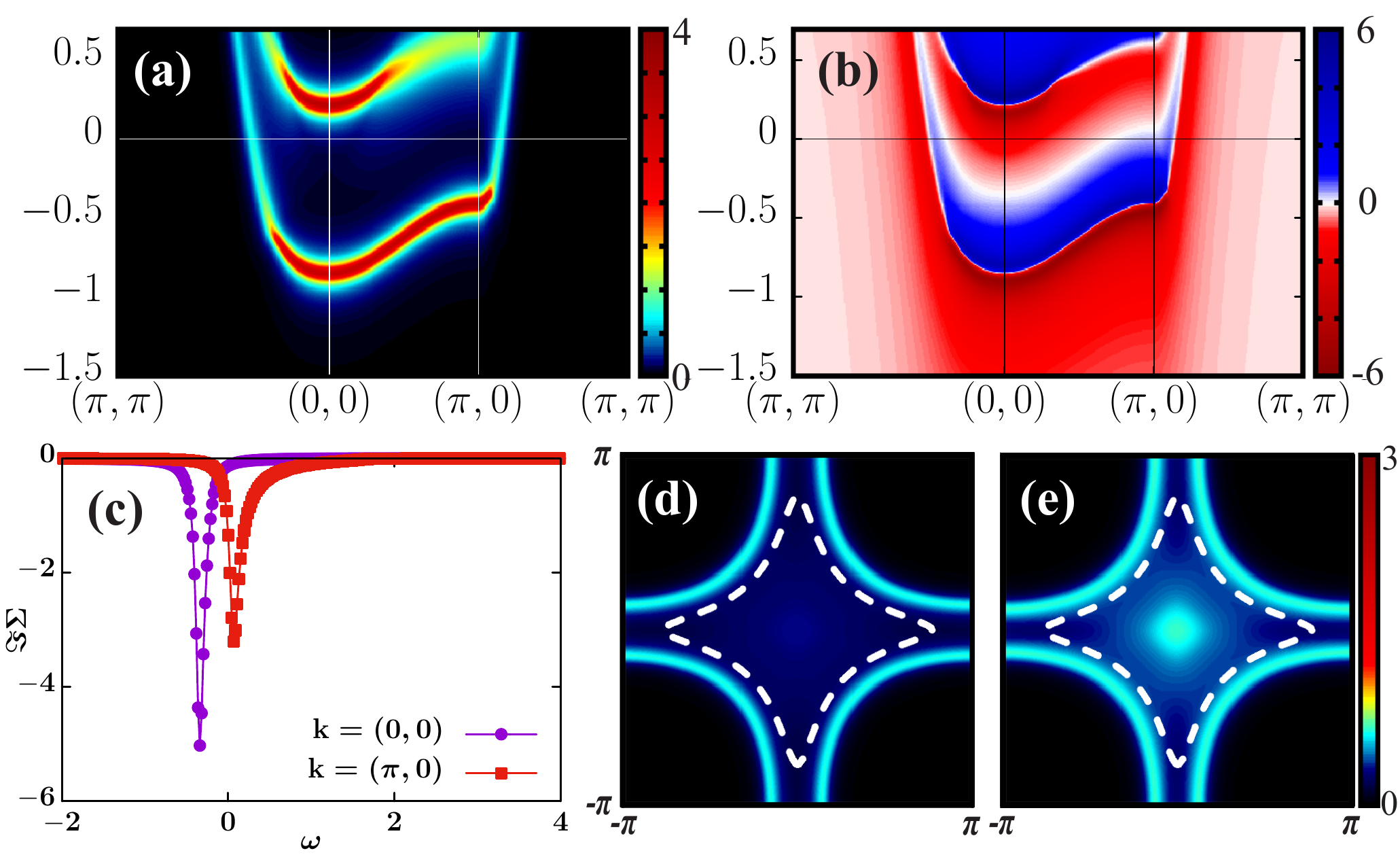}
        \caption{(a) Spectrum weight $A(\mathbf{k},\omega)$, (b) the real part of the Green's function, (c) the real frequency dependence of the imaginary part of the self-energy, and (d) the Fermi surface ($-\Im G(\mathbf{k},\omega_n=\pi/\beta)/\pi$) calculated by using the analytical expression in Eq.~(\ref{eq:analytical}) with $\xi=10,\  A\beta t^2=1.25,\ \delta=0.3\pi$ for $t^{\prime}=-0.45t$ and $n=0.40$. (e) The same plot as (d) but for $\xi=5$. Dashed curves denote the non-interacting Fermi surface.}
        \label{fig:analytical}
\end{figure}

In Fig.~\ref{fig:analytical}(a-c), we plot the Green's function and the self-energy using the analytical formula, Eq.~(\ref{eq:analytical}). We can see that the overall spectral and $G=0$ structures in D$\Gamma$A are well reproduced, while D$\Gamma$A results have a more complicated structure, likely due to other correlation effects (e.g., mass renormalization) not taken in Eq.~(\ref{eq:analytical}). We also note that the second term in Eq.~(\ref{eq:alpha}) is not relevant for the dispersive region, whereas it contributes to the partial flat band region where the Fermi velocity is small and makes the spectral weight asymmetric above and below the pseudogap. For example, our D$\Gamma$A result and analytical formula show that $\Im \Sigma(\omega)$ peak is slightly stronger at $\mathbf{k}=(0,0)$ than $\mathbf{k}=(\pi,0)$, which comes from the second term in Eq.~(\ref{eq:alpha}).

In Fig.~\ref{fig:analytical}(d,e), we also show the Fermi surface plot for $t^{\prime}=-0.45t$ and $n=0.40$, obtained with the analytical formula, Eq.~(\ref{eq:analytical}). The results reproduce the expanded Fermi surface structure seen in $U=8t$ results shown in Fig.~\ref{fig:fermisurface}. We also obtain a spectral weight around $\mathbf{k}=(0,0)$ for the smaller correlation length $\xi=5$, which is similar to the result of $U=6t$ at the same parameter. This consistency indicates that the observed Fermi surface expansion indeed originates from ferromagnetic fluctuations.

{\it Conclusion and Outlook.}---We investigated the effect of ferromagnetic fluctuations on the electronic spectrum within the two-dimensional Hubbard model without explicit ferromagnetic exchange interactions. We found that the Fermi surface gradually expands as the ferromagnetic fluctuations become more pronounced, accompanied by the emergence of a Luttinger surface and exchange splitting. Even with a moderate correlation length, the spectral weight changes dramatically in the partially flat band region. This finding suggests that similar phenomena may occur in real three-dimensional materials, despite the absence of strong two-dimensional fluctuations. We further obtained the analytical formulas that reproduce the spectral structures of our numerical results, which will be useful for analyzing realistic complex systems requiring a huge computational cost.

Since the ferromagnetic ($\mathbf{q}=(0,0)$) fluctuation mediates propagators with the same momentum, the curvature of the band at a certain momentum is the important measure of the correlation strength. This would also be the case for a system with such interactions, e.g., the Hatsugai-Kohmoto type interactions \cite{Hatsugai1992}.

It has also been pointed out that such a system shows unconventional large spatial pairing even without Fermi surface expansion in a weak to intermediate coupling regime \cite{Sayyad2020}. How the emergence of the Luttinger surface affects such exotic pairing in the strong coupling regime is an interesting direction for future work.

{\sl Acknowledgments}---We would like to thank Paul Worm and Karsten Held for fruitful discussions. This work was supported by JSPS KAKENHI Grant Numbers JP23H03817, JP24K17014, JP25K00961,  
JP23H04869, 
JP25H01506, JP25H01252
and 
JP23K03307, 
RIKEN TRIP initiative (RIKEN Quantum, Advanced General Intelligence for Science Program, Many-body Electron Systems),
and by MEXT as 
``Program for Promoting Researches on the Supercomputer Fugaku'' (Project ID: JPMXP1020230411).
We acknowledge Center for Computational Materials Science, Institute for Materials Research, Tohoku University for the use of MASAMUNE.

\bibliography{main}

\begin{thebibliography}{69}%
\makeatletter
\providecommand \@ifxundefined [1]{%
 \@ifx{#1\undefined}
}%
\providecommand \@ifnum [1]{%
 \ifnum #1\expandafter \@firstoftwo
 \else \expandafter \@secondoftwo
 \fi
}%
\providecommand \@ifx [1]{%
 \ifx #1\expandafter \@firstoftwo
 \else \expandafter \@secondoftwo
 \fi
}%
\providecommand \natexlab [1]{#1}%
\providecommand \enquote  [1]{``#1''}%
\providecommand \bibnamefont  [1]{#1}%
\providecommand \bibfnamefont [1]{#1}%
\providecommand \citenamefont [1]{#1}%
\providecommand \href@noop [0]{\@secondoftwo}%
\providecommand \href [0]{\begingroup \@sanitize@url \@href}%
\providecommand \@href[1]{\@@startlink{#1}\@@href}%
\providecommand \@@href[1]{\endgroup#1\@@endlink}%
\providecommand \@sanitize@url [0]{\catcode `\\12\catcode `\$12\catcode `\&12\catcode `\#12\catcode `\^12\catcode `\_12\catcode `\%12\relax}%
\providecommand \@@startlink[1]{}%
\providecommand \@@endlink[0]{}%
\providecommand \url  [0]{\begingroup\@sanitize@url \@url }%
\providecommand \@url [1]{\endgroup\@href {#1}{\urlprefix }}%
\providecommand \urlprefix  [0]{URL }%
\providecommand \Eprint [0]{\href }%
\providecommand \doibase [0]{http://dx.doi.org/}%
\providecommand \selectlanguage [0]{\@gobble}%
\providecommand \bibinfo  [0]{\@secondoftwo}%
\providecommand \bibfield  [0]{\@secondoftwo}%
\providecommand \translation [1]{[#1]}%
\providecommand \BibitemOpen [0]{}%
\providecommand \bibitemStop [0]{}%
\providecommand \bibitemNoStop [0]{.\EOS\space}%
\providecommand \EOS [0]{\spacefactor3000\relax}%
\providecommand \BibitemShut  [1]{\csname bibitem#1\endcsname}%
\let\auto@bib@innerbib\@empty
\bibitem [{\citenamefont {Momoi}\ and\ \citenamefont {Kubo}(1998)}]{Momoi1998}%
  \BibitemOpen
  \bibfield  {author} {\bibinfo {author} {\bibfnamefont {T.}~\bibnamefont {Momoi}}\ and\ \bibinfo {author} {\bibfnamefont {K.}~\bibnamefont {Kubo}},\ }\href {\doibase 10.1103/PhysRevB.58.R567} {\bibfield  {journal} {\bibinfo  {journal} {Phys. Rev. B}\ }\textbf {\bibinfo {volume} {58}},\ \bibinfo {pages} {R567} (\bibinfo {year} {1998})}\BibitemShut {NoStop}%
\bibitem [{\citenamefont {Held}\ and\ \citenamefont {Vollhardt}(1998)}]{Held1998}%
  \BibitemOpen
  \bibfield  {author} {\bibinfo {author} {\bibfnamefont {K.}~\bibnamefont {Held}}\ and\ \bibinfo {author} {\bibfnamefont {D.}~\bibnamefont {Vollhardt}},\ }\href@noop {} {\bibfield  {journal} {\bibinfo  {journal} {The European Physical Journal B-Condensed Matter and Complex Systems}\ }\textbf {\bibinfo {volume} {5}},\ \bibinfo {pages} {473} (\bibinfo {year} {1998})}\BibitemShut {NoStop}%
\bibitem [{\citenamefont {Lichtenstein}\ \emph {et~al.}(2001)\citenamefont {Lichtenstein}, \citenamefont {Katsnelson},\ and\ \citenamefont {Kotliar}}]{Lichtenstein2001}%
  \BibitemOpen
  \bibfield  {author} {\bibinfo {author} {\bibfnamefont {A.~I.}\ \bibnamefont {Lichtenstein}}, \bibinfo {author} {\bibfnamefont {M.~I.}\ \bibnamefont {Katsnelson}}, \ and\ \bibinfo {author} {\bibfnamefont {G.}~\bibnamefont {Kotliar}},\ }\href {\doibase 10.1103/PhysRevLett.87.067205} {\bibfield  {journal} {\bibinfo  {journal} {Phys. Rev. Lett.}\ }\textbf {\bibinfo {volume} {87}},\ \bibinfo {pages} {067205} (\bibinfo {year} {2001})}\BibitemShut {NoStop}%
\bibitem [{\citenamefont {Sakamoto}\ \emph {et~al.}(2002)\citenamefont {Sakamoto}, \citenamefont {Momoi},\ and\ \citenamefont {Kubo}}]{Sakamoto2002}%
  \BibitemOpen
  \bibfield  {author} {\bibinfo {author} {\bibfnamefont {H.}~\bibnamefont {Sakamoto}}, \bibinfo {author} {\bibfnamefont {T.}~\bibnamefont {Momoi}}, \ and\ \bibinfo {author} {\bibfnamefont {K.}~\bibnamefont {Kubo}},\ }\href {\doibase 10.1103/PhysRevB.65.224403} {\bibfield  {journal} {\bibinfo  {journal} {Phys. Rev. B}\ }\textbf {\bibinfo {volume} {65}},\ \bibinfo {pages} {224403} (\bibinfo {year} {2002})}\BibitemShut {NoStop}%
\bibitem [{\citenamefont {Sakai}\ \emph {et~al.}(2007)\citenamefont {Sakai}, \citenamefont {Arita},\ and\ \citenamefont {Aoki}}]{Sakai2007}%
  \BibitemOpen
  \bibfield  {author} {\bibinfo {author} {\bibfnamefont {S.}~\bibnamefont {Sakai}}, \bibinfo {author} {\bibfnamefont {R.}~\bibnamefont {Arita}}, \ and\ \bibinfo {author} {\bibfnamefont {H.}~\bibnamefont {Aoki}},\ }\href {\doibase 10.1103/PhysRevLett.99.216402} {\bibfield  {journal} {\bibinfo  {journal} {Phys. Rev. Lett.}\ }\textbf {\bibinfo {volume} {99}},\ \bibinfo {pages} {216402} (\bibinfo {year} {2007})}\BibitemShut {NoStop}%
\bibitem [{\citenamefont {Kanamori}(1963)}]{Kanamori1963}%
  \BibitemOpen
  \bibfield  {author} {\bibinfo {author} {\bibfnamefont {J.}~\bibnamefont {Kanamori}},\ }\href {\doibase 10.1143/PTP.30.275} {\bibfield  {journal} {\bibinfo  {journal} {Progress of Theoretical Physics}\ }\textbf {\bibinfo {volume} {30}},\ \bibinfo {pages} {275} (\bibinfo {year} {1963})}\BibitemShut {NoStop}%
\bibitem [{\citenamefont {Nagaoka}(1966)}]{Nagaoka1966}%
  \BibitemOpen
  \bibfield  {author} {\bibinfo {author} {\bibfnamefont {Y.}~\bibnamefont {Nagaoka}},\ }\href {\doibase 10.1103/PhysRev.147.392} {\bibfield  {journal} {\bibinfo  {journal} {Phys. Rev.}\ }\textbf {\bibinfo {volume} {147}},\ \bibinfo {pages} {392} (\bibinfo {year} {1966})}\BibitemShut {NoStop}%
\bibitem [{\citenamefont {Mielke}(1991)}]{Mielke1991}%
  \BibitemOpen
  \bibfield  {author} {\bibinfo {author} {\bibfnamefont {A.}~\bibnamefont {Mielke}},\ }\href {\doibase 10.1088/0305-4470/24/14/018} {\bibfield  {journal} {\bibinfo  {journal} {Journal of Physics A: Mathematical and General}\ }\textbf {\bibinfo {volume} {24}},\ \bibinfo {pages} {3311} (\bibinfo {year} {1991})}\BibitemShut {NoStop}%
\bibitem [{\citenamefont {Tasaki}(1992)}]{Tasaki1992}%
  \BibitemOpen
  \bibfield  {author} {\bibinfo {author} {\bibfnamefont {H.}~\bibnamefont {Tasaki}},\ }\href {\doibase 10.1103/PhysRevLett.69.1608} {\bibfield  {journal} {\bibinfo  {journal} {Phys. Rev. Lett.}\ }\textbf {\bibinfo {volume} {69}},\ \bibinfo {pages} {1608} (\bibinfo {year} {1992})}\BibitemShut {NoStop}%
\bibitem [{\citenamefont {Tasaki}(1995)}]{Tasaki1995}%
  \BibitemOpen
  \bibfield  {author} {\bibinfo {author} {\bibfnamefont {H.}~\bibnamefont {Tasaki}},\ }\href {\doibase 10.1103/PhysRevLett.75.4678} {\bibfield  {journal} {\bibinfo  {journal} {Phys. Rev. Lett.}\ }\textbf {\bibinfo {volume} {75}},\ \bibinfo {pages} {4678} (\bibinfo {year} {1995})}\BibitemShut {NoStop}%
\bibitem [{\citenamefont {Hlubina}\ \emph {et~al.}(1997)\citenamefont {Hlubina}, \citenamefont {Sorella},\ and\ \citenamefont {Guinea}}]{Hlubina1997}%
  \BibitemOpen
  \bibfield  {author} {\bibinfo {author} {\bibfnamefont {R.}~\bibnamefont {Hlubina}}, \bibinfo {author} {\bibfnamefont {S.}~\bibnamefont {Sorella}}, \ and\ \bibinfo {author} {\bibfnamefont {F.}~\bibnamefont {Guinea}},\ }\href {\doibase 10.1103/PhysRevLett.78.1343} {\bibfield  {journal} {\bibinfo  {journal} {Phys. Rev. Lett.}\ }\textbf {\bibinfo {volume} {78}},\ \bibinfo {pages} {1343} (\bibinfo {year} {1997})}\BibitemShut {NoStop}%
\bibitem [{\citenamefont {Hlubina}(1999)}]{Hlubina1999}%
  \BibitemOpen
  \bibfield  {author} {\bibinfo {author} {\bibfnamefont {R.}~\bibnamefont {Hlubina}},\ }\href {\doibase 10.1103/PhysRevB.59.9600} {\bibfield  {journal} {\bibinfo  {journal} {Phys. Rev. B}\ }\textbf {\bibinfo {volume} {59}},\ \bibinfo {pages} {9600} (\bibinfo {year} {1999})}\BibitemShut {NoStop}%
\bibitem [{\citenamefont {Irkhin}\ \emph {et~al.}(2001)\citenamefont {Irkhin}, \citenamefont {Katanin},\ and\ \citenamefont {Katsnelson}}]{Irkhin2001}%
  \BibitemOpen
  \bibfield  {author} {\bibinfo {author} {\bibfnamefont {V.~Y.}\ \bibnamefont {Irkhin}}, \bibinfo {author} {\bibfnamefont {A.~A.}\ \bibnamefont {Katanin}}, \ and\ \bibinfo {author} {\bibfnamefont {M.~I.}\ \bibnamefont {Katsnelson}},\ }\href {\doibase 10.1103/PhysRevB.64.165107} {\bibfield  {journal} {\bibinfo  {journal} {Phys. Rev. B}\ }\textbf {\bibinfo {volume} {64}},\ \bibinfo {pages} {165107} (\bibinfo {year} {2001})}\BibitemShut {NoStop}%
\bibitem [{\citenamefont {Honerkamp}\ and\ \citenamefont {Salmhofer}(2001{\natexlab{a}})}]{Honerkamp2001a}%
  \BibitemOpen
  \bibfield  {author} {\bibinfo {author} {\bibfnamefont {C.}~\bibnamefont {Honerkamp}}\ and\ \bibinfo {author} {\bibfnamefont {M.}~\bibnamefont {Salmhofer}},\ }\href {\doibase 10.1103/PhysRevLett.87.187004} {\bibfield  {journal} {\bibinfo  {journal} {Phys. Rev. Lett.}\ }\textbf {\bibinfo {volume} {87}},\ \bibinfo {pages} {187004} (\bibinfo {year} {2001}{\natexlab{a}})}\BibitemShut {NoStop}%
\bibitem [{\citenamefont {Honerkamp}\ and\ \citenamefont {Salmhofer}(2001{\natexlab{b}})}]{Honerkamp2001b}%
  \BibitemOpen
  \bibfield  {author} {\bibinfo {author} {\bibfnamefont {C.}~\bibnamefont {Honerkamp}}\ and\ \bibinfo {author} {\bibfnamefont {M.}~\bibnamefont {Salmhofer}},\ }\href {\doibase 10.1103/PhysRevB.64.184516} {\bibfield  {journal} {\bibinfo  {journal} {Phys. Rev. B}\ }\textbf {\bibinfo {volume} {64}},\ \bibinfo {pages} {184516} (\bibinfo {year} {2001}{\natexlab{b}})}\BibitemShut {NoStop}%
\bibitem [{\citenamefont {Kusakabe}\ and\ \citenamefont {Aoki}(1994{\natexlab{a}})}]{Kusakabe1994}%
  \BibitemOpen
  \bibfield  {author} {\bibinfo {author} {\bibfnamefont {K.}~\bibnamefont {Kusakabe}}\ and\ \bibinfo {author} {\bibfnamefont {H.}~\bibnamefont {Aoki}},\ }\href {\doibase 10.1103/PhysRevLett.72.144} {\bibfield  {journal} {\bibinfo  {journal} {Phys. Rev. Lett.}\ }\textbf {\bibinfo {volume} {72}},\ \bibinfo {pages} {144} (\bibinfo {year} {1994}{\natexlab{a}})}\BibitemShut {NoStop}%
\bibitem [{\citenamefont {Kusakabe}\ and\ \citenamefont {Aoki}(1994{\natexlab{b}})}]{Kusakabe1994b}%
  \BibitemOpen
  \bibfield  {author} {\bibinfo {author} {\bibfnamefont {K.}~\bibnamefont {Kusakabe}}\ and\ \bibinfo {author} {\bibfnamefont {H.}~\bibnamefont {Aoki}},\ }\href {\doibase 10.1103/PhysRevB.50.12991} {\bibfield  {journal} {\bibinfo  {journal} {Phys. Rev. B}\ }\textbf {\bibinfo {volume} {50}},\ \bibinfo {pages} {12991} (\bibinfo {year} {1994}{\natexlab{b}})}\BibitemShut {NoStop}%
\bibitem [{\citenamefont {Pickel}\ \emph {et~al.}(2010)\citenamefont {Pickel}, \citenamefont {Schmidt}, \citenamefont {Weinelt},\ and\ \citenamefont {Donath}}]{Pickel2010}%
  \BibitemOpen
  \bibfield  {author} {\bibinfo {author} {\bibfnamefont {M.}~\bibnamefont {Pickel}}, \bibinfo {author} {\bibfnamefont {A.~B.}\ \bibnamefont {Schmidt}}, \bibinfo {author} {\bibfnamefont {M.}~\bibnamefont {Weinelt}}, \ and\ \bibinfo {author} {\bibfnamefont {M.}~\bibnamefont {Donath}},\ }\href {\doibase 10.1103/PhysRevLett.104.237204} {\bibfield  {journal} {\bibinfo  {journal} {Phys. Rev. Lett.}\ }\textbf {\bibinfo {volume} {104}},\ \bibinfo {pages} {237204} (\bibinfo {year} {2010})}\BibitemShut {NoStop}%
\bibitem [{\citenamefont {Jeong}\ \emph {et~al.}(2013)\citenamefont {Jeong}, \citenamefont {Choi}, \citenamefont {Kim}, \citenamefont {Chang}, \citenamefont {Sohn}, \citenamefont {Park}, \citenamefont {Kang}, \citenamefont {Cho}, \citenamefont {Baek}, \citenamefont {Eom}, \citenamefont {Shim}, \citenamefont {Yu}, \citenamefont {Kim}, \citenamefont {Moon},\ and\ \citenamefont {Noh}}]{Jeong2013}%
  \BibitemOpen
  \bibfield  {author} {\bibinfo {author} {\bibfnamefont {D.~W.}\ \bibnamefont {Jeong}}, \bibinfo {author} {\bibfnamefont {H.~C.}\ \bibnamefont {Choi}}, \bibinfo {author} {\bibfnamefont {C.~H.}\ \bibnamefont {Kim}}, \bibinfo {author} {\bibfnamefont {S.~H.}\ \bibnamefont {Chang}}, \bibinfo {author} {\bibfnamefont {C.~H.}\ \bibnamefont {Sohn}}, \bibinfo {author} {\bibfnamefont {H.~J.}\ \bibnamefont {Park}}, \bibinfo {author} {\bibfnamefont {T.~D.}\ \bibnamefont {Kang}}, \bibinfo {author} {\bibfnamefont {D.-Y.}\ \bibnamefont {Cho}}, \bibinfo {author} {\bibfnamefont {S.~H.}\ \bibnamefont {Baek}}, \bibinfo {author} {\bibfnamefont {C.~B.}\ \bibnamefont {Eom}}, \bibinfo {author} {\bibfnamefont {J.~H.}\ \bibnamefont {Shim}}, \bibinfo {author} {\bibfnamefont {J.}~\bibnamefont {Yu}}, \bibinfo {author} {\bibfnamefont {K.~W.}\ \bibnamefont {Kim}}, \bibinfo {author} {\bibfnamefont {S.~J.}\ \bibnamefont {Moon}}, \ and\ \bibinfo {author} {\bibfnamefont {T.~W.}\ \bibnamefont {Noh}},\ }\href {\doibase
  10.1103/PhysRevLett.110.247202} {\bibfield  {journal} {\bibinfo  {journal} {Phys. Rev. Lett.}\ }\textbf {\bibinfo {volume} {110}},\ \bibinfo {pages} {247202} (\bibinfo {year} {2013})}\BibitemShut {NoStop}%
\bibitem [{\citenamefont {Hahn}\ \emph {et~al.}(2021)\citenamefont {Hahn}, \citenamefont {Sohn}, \citenamefont {Kim}, \citenamefont {Kim}, \citenamefont {Huh}, \citenamefont {Kim}, \citenamefont {Kyung}, \citenamefont {Kim}, \citenamefont {Kim}, \citenamefont {Kim}, \citenamefont {Noh}, \citenamefont {Shim},\ and\ \citenamefont {Kim}}]{Hahn2021}%
  \BibitemOpen
  \bibfield  {author} {\bibinfo {author} {\bibfnamefont {S.}~\bibnamefont {Hahn}}, \bibinfo {author} {\bibfnamefont {B.}~\bibnamefont {Sohn}}, \bibinfo {author} {\bibfnamefont {M.}~\bibnamefont {Kim}}, \bibinfo {author} {\bibfnamefont {J.~R.}\ \bibnamefont {Kim}}, \bibinfo {author} {\bibfnamefont {S.}~\bibnamefont {Huh}}, \bibinfo {author} {\bibfnamefont {Y.}~\bibnamefont {Kim}}, \bibinfo {author} {\bibfnamefont {W.}~\bibnamefont {Kyung}}, \bibinfo {author} {\bibfnamefont {M.}~\bibnamefont {Kim}}, \bibinfo {author} {\bibfnamefont {D.}~\bibnamefont {Kim}}, \bibinfo {author} {\bibfnamefont {Y.}~\bibnamefont {Kim}}, \bibinfo {author} {\bibfnamefont {T.~W.}\ \bibnamefont {Noh}}, \bibinfo {author} {\bibfnamefont {J.~H.}\ \bibnamefont {Shim}}, \ and\ \bibinfo {author} {\bibfnamefont {C.}~\bibnamefont {Kim}},\ }\href {\doibase 10.1103/PhysRevLett.127.256401} {\bibfield  {journal} {\bibinfo  {journal} {Phys. Rev. Lett.}\ }\textbf {\bibinfo {volume} {127}},\ \bibinfo {pages} {256401} (\bibinfo {year}
  {2021})}\BibitemShut {NoStop}%
\bibitem [{\citenamefont {Eich}\ \emph {et~al.}(2017)\citenamefont {Eich}, \citenamefont {Plötzing}, \citenamefont {Rollinger}, \citenamefont {Emmerich}, \citenamefont {Adam}, \citenamefont {Chen}, \citenamefont {Kapteyn}, \citenamefont {Murnane}, \citenamefont {Plucinski}, \citenamefont {Steil}, \citenamefont {Stadtmüller}, \citenamefont {Cinchetti}, \citenamefont {Aeschlimann}, \citenamefont {Schneider},\ and\ \citenamefont {Mathias}}]{Eich2017}%
  \BibitemOpen
  \bibfield  {author} {\bibinfo {author} {\bibfnamefont {S.}~\bibnamefont {Eich}}, \bibinfo {author} {\bibfnamefont {M.}~\bibnamefont {Plötzing}}, \bibinfo {author} {\bibfnamefont {M.}~\bibnamefont {Rollinger}}, \bibinfo {author} {\bibfnamefont {S.}~\bibnamefont {Emmerich}}, \bibinfo {author} {\bibfnamefont {R.}~\bibnamefont {Adam}}, \bibinfo {author} {\bibfnamefont {C.}~\bibnamefont {Chen}}, \bibinfo {author} {\bibfnamefont {H.~C.}\ \bibnamefont {Kapteyn}}, \bibinfo {author} {\bibfnamefont {M.~M.}\ \bibnamefont {Murnane}}, \bibinfo {author} {\bibfnamefont {L.}~\bibnamefont {Plucinski}}, \bibinfo {author} {\bibfnamefont {D.}~\bibnamefont {Steil}}, \bibinfo {author} {\bibfnamefont {B.}~\bibnamefont {Stadtmüller}}, \bibinfo {author} {\bibfnamefont {M.}~\bibnamefont {Cinchetti}}, \bibinfo {author} {\bibfnamefont {M.}~\bibnamefont {Aeschlimann}}, \bibinfo {author} {\bibfnamefont {C.~M.}\ \bibnamefont {Schneider}}, \ and\ \bibinfo {author} {\bibfnamefont {S.}~\bibnamefont {Mathias}},\ }\href {\doibase
  10.1126/sciadv.1602094} {\bibfield  {journal} {\bibinfo  {journal} {Science Advances}\ }\textbf {\bibinfo {volume} {3}},\ \bibinfo {pages} {e1602094} (\bibinfo {year} {2017})},\ \Eprint {http://arxiv.org/abs/https://www.science.org/doi/pdf/10.1126/sciadv.1602094} {https://www.science.org/doi/pdf/10.1126/sciadv.1602094} \BibitemShut {NoStop}%
\bibitem [{\citenamefont {Monthoux}(2003)}]{Monthoux2003}%
  \BibitemOpen
  \bibfield  {author} {\bibinfo {author} {\bibfnamefont {P.}~\bibnamefont {Monthoux}},\ }\href {\doibase 10.1103/PhysRevB.68.064408} {\bibfield  {journal} {\bibinfo  {journal} {Phys. Rev. B}\ }\textbf {\bibinfo {volume} {68}},\ \bibinfo {pages} {064408} (\bibinfo {year} {2003})}\BibitemShut {NoStop}%
\bibitem [{\citenamefont {Hankevych}\ \emph {et~al.}(2003)\citenamefont {Hankevych}, \citenamefont {Kyung},\ and\ \citenamefont {Tremblay}}]{Hankevych2003}%
  \BibitemOpen
  \bibfield  {author} {\bibinfo {author} {\bibfnamefont {V.}~\bibnamefont {Hankevych}}, \bibinfo {author} {\bibfnamefont {B.}~\bibnamefont {Kyung}}, \ and\ \bibinfo {author} {\bibfnamefont {A.-M.~S.}\ \bibnamefont {Tremblay}},\ }\href {\doibase 10.1103/PhysRevB.68.214405} {\bibfield  {journal} {\bibinfo  {journal} {Phys. Rev. B}\ }\textbf {\bibinfo {volume} {68}},\ \bibinfo {pages} {214405} (\bibinfo {year} {2003})}\BibitemShut {NoStop}%
\bibitem [{\citenamefont {Katanin}\ \emph {et~al.}(2005)\citenamefont {Katanin}, \citenamefont {Kampf},\ and\ \citenamefont {Irkhin}}]{Katanin2005}%
  \BibitemOpen
  \bibfield  {author} {\bibinfo {author} {\bibfnamefont {A.~A.}\ \bibnamefont {Katanin}}, \bibinfo {author} {\bibfnamefont {A.~P.}\ \bibnamefont {Kampf}}, \ and\ \bibinfo {author} {\bibfnamefont {V.~Y.}\ \bibnamefont {Irkhin}},\ }\href {\doibase 10.1103/PhysRevB.71.085105} {\bibfield  {journal} {\bibinfo  {journal} {Phys. Rev. B}\ }\textbf {\bibinfo {volume} {71}},\ \bibinfo {pages} {085105} (\bibinfo {year} {2005})}\BibitemShut {NoStop}%
\bibitem [{\citenamefont {Katanin}(2005)}]{Katanin2005b}%
  \BibitemOpen
  \bibfield  {author} {\bibinfo {author} {\bibfnamefont {A.~A.}\ \bibnamefont {Katanin}},\ }\href {\doibase 10.1103/PhysRevB.72.035111} {\bibfield  {journal} {\bibinfo  {journal} {Phys. Rev. B}\ }\textbf {\bibinfo {volume} {72}},\ \bibinfo {pages} {035111} (\bibinfo {year} {2005})}\BibitemShut {NoStop}%
\bibitem [{\citenamefont {Katanin}\ and\ \citenamefont {Irkhin}(2008)}]{Katanin2008}%
  \BibitemOpen
  \bibfield  {author} {\bibinfo {author} {\bibfnamefont {A.~A.}\ \bibnamefont {Katanin}}\ and\ \bibinfo {author} {\bibfnamefont {V.~Y.}\ \bibnamefont {Irkhin}},\ }\href {\doibase 10.1103/PhysRevB.77.115129} {\bibfield  {journal} {\bibinfo  {journal} {Phys. Rev. B}\ }\textbf {\bibinfo {volume} {77}},\ \bibinfo {pages} {115129} (\bibinfo {year} {2008})}\BibitemShut {NoStop}%
\bibitem [{\citenamefont {Sayyad}\ \emph {et~al.}(2020)\citenamefont {Sayyad}, \citenamefont {Huang}, \citenamefont {Kitatani}, \citenamefont {Vaezi}, \citenamefont {Nussinov}, \citenamefont {Vaezi},\ and\ \citenamefont {Aoki}}]{Sayyad2020}%
  \BibitemOpen
  \bibfield  {author} {\bibinfo {author} {\bibfnamefont {S.}~\bibnamefont {Sayyad}}, \bibinfo {author} {\bibfnamefont {E.~W.}\ \bibnamefont {Huang}}, \bibinfo {author} {\bibfnamefont {M.}~\bibnamefont {Kitatani}}, \bibinfo {author} {\bibfnamefont {M.-S.}\ \bibnamefont {Vaezi}}, \bibinfo {author} {\bibfnamefont {Z.}~\bibnamefont {Nussinov}}, \bibinfo {author} {\bibfnamefont {A.}~\bibnamefont {Vaezi}}, \ and\ \bibinfo {author} {\bibfnamefont {H.}~\bibnamefont {Aoki}},\ }\href {\doibase 10.1103/PhysRevB.101.014501} {\bibfield  {journal} {\bibinfo  {journal} {Phys. Rev. B}\ }\textbf {\bibinfo {volume} {101}},\ \bibinfo {pages} {014501} (\bibinfo {year} {2020})}\BibitemShut {NoStop}%
\bibitem [{\citenamefont {Sayyad}\ \emph {et~al.}(2023)\citenamefont {Sayyad}, \citenamefont {Kitatani}, \citenamefont {Vaezi},\ and\ \citenamefont {Aoki}}]{Sayyad2023}%
  \BibitemOpen
  \bibfield  {author} {\bibinfo {author} {\bibfnamefont {S.}~\bibnamefont {Sayyad}}, \bibinfo {author} {\bibfnamefont {M.}~\bibnamefont {Kitatani}}, \bibinfo {author} {\bibfnamefont {A.}~\bibnamefont {Vaezi}}, \ and\ \bibinfo {author} {\bibfnamefont {H.}~\bibnamefont {Aoki}},\ }\href@noop {} {\bibfield  {journal} {\bibinfo  {journal} {Journal of Physics: Condensed Matter}\ }\textbf {\bibinfo {volume} {35}},\ \bibinfo {pages} {245605} (\bibinfo {year} {2023})}\BibitemShut {NoStop}%
\bibitem [{\citenamefont {Lichtenstein}\ and\ \citenamefont {Katsnelson}(2000)}]{Lichtenstein2000}%
  \BibitemOpen
  \bibfield  {author} {\bibinfo {author} {\bibfnamefont {A.~I.}\ \bibnamefont {Lichtenstein}}\ and\ \bibinfo {author} {\bibfnamefont {M.~I.}\ \bibnamefont {Katsnelson}},\ }\href {\doibase 10.1103/PhysRevB.62.R9283} {\bibfield  {journal} {\bibinfo  {journal} {Phys. Rev. B}\ }\textbf {\bibinfo {volume} {62}},\ \bibinfo {pages} {R9283} (\bibinfo {year} {2000})}\BibitemShut {NoStop}%
\bibitem [{\citenamefont {Kotliar}\ \emph {et~al.}(2001)\citenamefont {Kotliar}, \citenamefont {Savrasov}, \citenamefont {P\'alsson},\ and\ \citenamefont {Biroli}}]{Kotliar2001}%
  \BibitemOpen
  \bibfield  {author} {\bibinfo {author} {\bibfnamefont {G.}~\bibnamefont {Kotliar}}, \bibinfo {author} {\bibfnamefont {S.~Y.}\ \bibnamefont {Savrasov}}, \bibinfo {author} {\bibfnamefont {G.}~\bibnamefont {P\'alsson}}, \ and\ \bibinfo {author} {\bibfnamefont {G.}~\bibnamefont {Biroli}},\ }\href {\doibase 10.1103/PhysRevLett.87.186401} {\bibfield  {journal} {\bibinfo  {journal} {Phys. Rev. Lett.}\ }\textbf {\bibinfo {volume} {87}},\ \bibinfo {pages} {186401} (\bibinfo {year} {2001})}\BibitemShut {NoStop}%
\bibitem [{\citenamefont {Maier}\ \emph {et~al.}(2005)\citenamefont {Maier}, \citenamefont {Jarrell}, \citenamefont {Pruschke},\ and\ \citenamefont {Hettler}}]{Maier2005}%
  \BibitemOpen
  \bibfield  {author} {\bibinfo {author} {\bibfnamefont {T.}~\bibnamefont {Maier}}, \bibinfo {author} {\bibfnamefont {M.}~\bibnamefont {Jarrell}}, \bibinfo {author} {\bibfnamefont {T.}~\bibnamefont {Pruschke}}, \ and\ \bibinfo {author} {\bibfnamefont {M.~H.}\ \bibnamefont {Hettler}},\ }\href {\doibase 10.1103/RevModPhys.77.1027} {\bibfield  {journal} {\bibinfo  {journal} {Rev. Mod. Phys.}\ }\textbf {\bibinfo {volume} {77}},\ \bibinfo {pages} {1027} (\bibinfo {year} {2005})}\BibitemShut {NoStop}%
\bibitem [{\citenamefont {Metzner}\ and\ \citenamefont {Vollhardt}(1989)}]{Metzner1989}%
  \BibitemOpen
  \bibfield  {author} {\bibinfo {author} {\bibfnamefont {W.}~\bibnamefont {Metzner}}\ and\ \bibinfo {author} {\bibfnamefont {D.}~\bibnamefont {Vollhardt}},\ }\href {\doibase 10.1103/PhysRevLett.62.324} {\bibfield  {journal} {\bibinfo  {journal} {Phys. Rev. Lett.}\ }\textbf {\bibinfo {volume} {62}},\ \bibinfo {pages} {324} (\bibinfo {year} {1989})}\BibitemShut {NoStop}%
\bibitem [{\citenamefont {Georges}\ and\ \citenamefont {Kotliar}(1992)}]{Georges1992}%
  \BibitemOpen
  \bibfield  {author} {\bibinfo {author} {\bibfnamefont {A.}~\bibnamefont {Georges}}\ and\ \bibinfo {author} {\bibfnamefont {G.}~\bibnamefont {Kotliar}},\ }\href {\doibase 10.1103/PhysRevB.45.6479} {\bibfield  {journal} {\bibinfo  {journal} {Phys. Rev. B}\ }\textbf {\bibinfo {volume} {45}},\ \bibinfo {pages} {6479} (\bibinfo {year} {1992})}\BibitemShut {NoStop}%
\bibitem [{\citenamefont {Georges}\ \emph {et~al.}(1996)\citenamefont {Georges}, \citenamefont {Kotliar}, \citenamefont {Krauth},\ and\ \citenamefont {Rozenberg}}]{Georges1996}%
  \BibitemOpen
  \bibfield  {author} {\bibinfo {author} {\bibfnamefont {A.}~\bibnamefont {Georges}}, \bibinfo {author} {\bibfnamefont {G.}~\bibnamefont {Kotliar}}, \bibinfo {author} {\bibfnamefont {W.}~\bibnamefont {Krauth}}, \ and\ \bibinfo {author} {\bibfnamefont {M.~J.}\ \bibnamefont {Rozenberg}},\ }\href {\doibase 10.1103/RevModPhys.68.13} {\bibfield  {journal} {\bibinfo  {journal} {Rev. Mod. Phys.}\ }\textbf {\bibinfo {volume} {68}},\ \bibinfo {pages} {13} (\bibinfo {year} {1996})}\BibitemShut {NoStop}%
\bibitem [{\citenamefont {Nomura}\ \emph {et~al.}(2015)\citenamefont {Nomura}, \citenamefont {Sakai},\ and\ \citenamefont {Arita}}]{Nomura2015}%
  \BibitemOpen
  \bibfield  {author} {\bibinfo {author} {\bibfnamefont {Y.}~\bibnamefont {Nomura}}, \bibinfo {author} {\bibfnamefont {S.}~\bibnamefont {Sakai}}, \ and\ \bibinfo {author} {\bibfnamefont {R.}~\bibnamefont {Arita}},\ }\href {\doibase 10.1103/PhysRevB.91.235107} {\bibfield  {journal} {\bibinfo  {journal} {Phys. Rev. B}\ }\textbf {\bibinfo {volume} {91}},\ \bibinfo {pages} {235107} (\bibinfo {year} {2015})}\BibitemShut {NoStop}%
\bibitem [{\citenamefont {Nomura}\ \emph {et~al.}(2022)\citenamefont {Nomura}, \citenamefont {Sakai},\ and\ \citenamefont {Arita}}]{Nomura2022}%
  \BibitemOpen
  \bibfield  {author} {\bibinfo {author} {\bibfnamefont {Y.}~\bibnamefont {Nomura}}, \bibinfo {author} {\bibfnamefont {S.}~\bibnamefont {Sakai}}, \ and\ \bibinfo {author} {\bibfnamefont {R.}~\bibnamefont {Arita}},\ }\href {\doibase 10.1103/PhysRevLett.128.206401} {\bibfield  {journal} {\bibinfo  {journal} {Phys. Rev. Lett.}\ }\textbf {\bibinfo {volume} {128}},\ \bibinfo {pages} {206401} (\bibinfo {year} {2022})}\BibitemShut {NoStop}%
\bibitem [{\citenamefont {Keimer}\ \emph {et~al.}(2015)\citenamefont {Keimer}, \citenamefont {Kivelson}, \citenamefont {Norman}, \citenamefont {Uchida},\ and\ \citenamefont {Zaanen}}]{Keimer2015}%
  \BibitemOpen
  \bibfield  {author} {\bibinfo {author} {\bibfnamefont {B.}~\bibnamefont {Keimer}}, \bibinfo {author} {\bibfnamefont {S.~A.}\ \bibnamefont {Kivelson}}, \bibinfo {author} {\bibfnamefont {M.~R.}\ \bibnamefont {Norman}}, \bibinfo {author} {\bibfnamefont {S.}~\bibnamefont {Uchida}}, \ and\ \bibinfo {author} {\bibfnamefont {J.}~\bibnamefont {Zaanen}},\ }\href {https://www.nature.com/articles/nature14165} {\bibfield  {journal} {\bibinfo  {journal} {Nature}\ }\textbf {\bibinfo {volume} {518}},\ \bibinfo {pages} {179} (\bibinfo {year} {2015})}\BibitemShut {NoStop}%
\bibitem [{\citenamefont {Vilk}\ and\ \citenamefont {Tremblay}(1997)}]{Vilk1997}%
  \BibitemOpen
  \bibfield  {author} {\bibinfo {author} {\bibfnamefont {Y.}~\bibnamefont {Vilk}}\ and\ \bibinfo {author} {\bibfnamefont {A.-M.}\ \bibnamefont {Tremblay}},\ }\href@noop {} {\bibfield  {journal} {\bibinfo  {journal} {Journal de Physique I}\ }\textbf {\bibinfo {volume} {7}},\ \bibinfo {pages} {1309} (\bibinfo {year} {1997})}\BibitemShut {NoStop}%
\bibitem [{\citenamefont {Borejsza}\ and\ \citenamefont {Dupuis}(2004)}]{Borejsza2004}%
  \BibitemOpen
  \bibfield  {author} {\bibinfo {author} {\bibfnamefont {K.}~\bibnamefont {Borejsza}}\ and\ \bibinfo {author} {\bibfnamefont {N.}~\bibnamefont {Dupuis}},\ }\href {\doibase 10.1103/PhysRevB.69.085119} {\bibfield  {journal} {\bibinfo  {journal} {Phys. Rev. B}\ }\textbf {\bibinfo {volume} {69}},\ \bibinfo {pages} {085119} (\bibinfo {year} {2004})}\BibitemShut {NoStop}%
\bibitem [{\citenamefont {Sch\"afer}\ \emph {et~al.}(2021)\citenamefont {Sch\"afer}, \citenamefont {Wentzell}, \citenamefont {\ifmmode~\check{S}\else \v{S}\fi{}imkovic}, \citenamefont {He}, \citenamefont {Hille}, \citenamefont {Klett}, \citenamefont {Eckhardt}, \citenamefont {Arzhang}, \citenamefont {Harkov}, \citenamefont {Le~R\'egent}, \citenamefont {Kirsch}, \citenamefont {Wang}, \citenamefont {Kim}, \citenamefont {Kozik}, \citenamefont {Stepanov}, \citenamefont {Kauch}, \citenamefont {Andergassen}, \citenamefont {Hansmann}, \citenamefont {Rohe}, \citenamefont {Vilk}, \citenamefont {LeBlanc}, \citenamefont {Zhang}, \citenamefont {Tremblay}, \citenamefont {Ferrero}, \citenamefont {Parcollet},\ and\ \citenamefont {Georges}}]{Schaefer2021}%
  \BibitemOpen
  \bibfield  {author} {\bibinfo {author} {\bibfnamefont {T.}~\bibnamefont {Sch\"afer}}, \bibinfo {author} {\bibfnamefont {N.}~\bibnamefont {Wentzell}}, \bibinfo {author} {\bibfnamefont {F.}~\bibnamefont {\ifmmode~\check{S}\else \v{S}\fi{}imkovic}}, \bibinfo {author} {\bibfnamefont {Y.-Y.}\ \bibnamefont {He}}, \bibinfo {author} {\bibfnamefont {C.}~\bibnamefont {Hille}}, \bibinfo {author} {\bibfnamefont {M.}~\bibnamefont {Klett}}, \bibinfo {author} {\bibfnamefont {C.~J.}\ \bibnamefont {Eckhardt}}, \bibinfo {author} {\bibfnamefont {B.}~\bibnamefont {Arzhang}}, \bibinfo {author} {\bibfnamefont {V.}~\bibnamefont {Harkov}}, \bibinfo {author} {\bibfnamefont {F.~m. c.-M.}\ \bibnamefont {Le~R\'egent}}, \bibinfo {author} {\bibfnamefont {A.}~\bibnamefont {Kirsch}}, \bibinfo {author} {\bibfnamefont {Y.}~\bibnamefont {Wang}}, \bibinfo {author} {\bibfnamefont {A.~J.}\ \bibnamefont {Kim}}, \bibinfo {author} {\bibfnamefont {E.}~\bibnamefont {Kozik}}, \bibinfo {author} {\bibfnamefont {E.~A.}\ \bibnamefont {Stepanov}},
  \bibinfo {author} {\bibfnamefont {A.}~\bibnamefont {Kauch}}, \bibinfo {author} {\bibfnamefont {S.}~\bibnamefont {Andergassen}}, \bibinfo {author} {\bibfnamefont {P.}~\bibnamefont {Hansmann}}, \bibinfo {author} {\bibfnamefont {D.}~\bibnamefont {Rohe}}, \bibinfo {author} {\bibfnamefont {Y.~M.}\ \bibnamefont {Vilk}}, \bibinfo {author} {\bibfnamefont {J.~P.~F.}\ \bibnamefont {LeBlanc}}, \bibinfo {author} {\bibfnamefont {S.}~\bibnamefont {Zhang}}, \bibinfo {author} {\bibfnamefont {A.-M.~S.}\ \bibnamefont {Tremblay}}, \bibinfo {author} {\bibfnamefont {M.}~\bibnamefont {Ferrero}}, \bibinfo {author} {\bibfnamefont {O.}~\bibnamefont {Parcollet}}, \ and\ \bibinfo {author} {\bibfnamefont {A.}~\bibnamefont {Georges}},\ }\href {\doibase 10.1103/PhysRevX.11.011058} {\bibfield  {journal} {\bibinfo  {journal} {Phys. Rev. X}\ }\textbf {\bibinfo {volume} {11}},\ \bibinfo {pages} {011058} (\bibinfo {year} {2021})}\BibitemShut {NoStop}%
\bibitem [{\citenamefont {Schmalian}\ \emph {et~al.}(1999)\citenamefont {Schmalian}, \citenamefont {Pines},\ and\ \citenamefont {Stojkovi\ifmmode~\acute{c}\else \'{c}\fi{}}}]{Schmalian1999}%
  \BibitemOpen
  \bibfield  {author} {\bibinfo {author} {\bibfnamefont {J.}~\bibnamefont {Schmalian}}, \bibinfo {author} {\bibfnamefont {D.}~\bibnamefont {Pines}}, \ and\ \bibinfo {author} {\bibfnamefont {B.}~\bibnamefont {Stojkovi\ifmmode~\acute{c}\else \'{c}\fi{}}},\ }\href {\doibase 10.1103/PhysRevB.60.667} {\bibfield  {journal} {\bibinfo  {journal} {Phys. Rev. B}\ }\textbf {\bibinfo {volume} {60}},\ \bibinfo {pages} {667} (\bibinfo {year} {1999})}\BibitemShut {NoStop}%
\bibitem [{\citenamefont {Sch\"afer}\ \emph {et~al.}(2015)\citenamefont {Sch\"afer}, \citenamefont {Geles}, \citenamefont {Rost}, \citenamefont {Rohringer}, \citenamefont {Arrigoni}, \citenamefont {Held}, \citenamefont {Bl\"umer}, \citenamefont {Aichhorn},\ and\ \citenamefont {Toschi}}]{Schaefer2015}%
  \BibitemOpen
  \bibfield  {author} {\bibinfo {author} {\bibfnamefont {T.}~\bibnamefont {Sch\"afer}}, \bibinfo {author} {\bibfnamefont {F.}~\bibnamefont {Geles}}, \bibinfo {author} {\bibfnamefont {D.}~\bibnamefont {Rost}}, \bibinfo {author} {\bibfnamefont {G.}~\bibnamefont {Rohringer}}, \bibinfo {author} {\bibfnamefont {E.}~\bibnamefont {Arrigoni}}, \bibinfo {author} {\bibfnamefont {K.}~\bibnamefont {Held}}, \bibinfo {author} {\bibfnamefont {N.}~\bibnamefont {Bl\"umer}}, \bibinfo {author} {\bibfnamefont {M.}~\bibnamefont {Aichhorn}}, \ and\ \bibinfo {author} {\bibfnamefont {A.}~\bibnamefont {Toschi}},\ }\href {\doibase 10.1103/PhysRevB.91.125109} {\bibfield  {journal} {\bibinfo  {journal} {Phys. Rev. B}\ }\textbf {\bibinfo {volume} {91}},\ \bibinfo {pages} {125109} (\bibinfo {year} {2015})}\BibitemShut {NoStop}%
\bibitem [{\citenamefont {Krien}\ \emph {et~al.}(2022)\citenamefont {Krien}, \citenamefont {Worm}, \citenamefont {Chalupa-Gantner}, \citenamefont {Toschi},\ and\ \citenamefont {Held}}]{Krien2022}%
  \BibitemOpen
  \bibfield  {author} {\bibinfo {author} {\bibfnamefont {F.}~\bibnamefont {Krien}}, \bibinfo {author} {\bibfnamefont {P.}~\bibnamefont {Worm}}, \bibinfo {author} {\bibfnamefont {P.}~\bibnamefont {Chalupa-Gantner}}, \bibinfo {author} {\bibfnamefont {A.}~\bibnamefont {Toschi}}, \ and\ \bibinfo {author} {\bibfnamefont {K.}~\bibnamefont {Held}},\ }\href@noop {} {\bibfield  {journal} {\bibinfo  {journal} {Communications Physics}\ }\textbf {\bibinfo {volume} {5}},\ \bibinfo {pages} {336} (\bibinfo {year} {2022})}\BibitemShut {NoStop}%
\bibitem [{\citenamefont {Toschi}\ \emph {et~al.}(2007)\citenamefont {Toschi}, \citenamefont {Katanin},\ and\ \citenamefont {Held}}]{Toschi2007}%
  \BibitemOpen
  \bibfield  {author} {\bibinfo {author} {\bibfnamefont {A.}~\bibnamefont {Toschi}}, \bibinfo {author} {\bibfnamefont {A.~A.}\ \bibnamefont {Katanin}}, \ and\ \bibinfo {author} {\bibfnamefont {K.}~\bibnamefont {Held}},\ }\href {\doibase 10.1103/PhysRevB.75.045118} {\bibfield  {journal} {\bibinfo  {journal} {Phys. Rev. B}\ }\textbf {\bibinfo {volume} {75}},\ \bibinfo {pages} {045118} (\bibinfo {year} {2007})}\BibitemShut {NoStop}%
\bibitem [{\citenamefont {Katanin}\ \emph {et~al.}(2009)\citenamefont {Katanin}, \citenamefont {Toschi},\ and\ \citenamefont {Held}}]{Katanin2009}%
  \BibitemOpen
  \bibfield  {author} {\bibinfo {author} {\bibfnamefont {A.~A.}\ \bibnamefont {Katanin}}, \bibinfo {author} {\bibfnamefont {A.}~\bibnamefont {Toschi}}, \ and\ \bibinfo {author} {\bibfnamefont {K.}~\bibnamefont {Held}},\ }\href {\doibase 10.1103/PhysRevB.80.075104} {\bibfield  {journal} {\bibinfo  {journal} {Phys. Rev. B}\ }\textbf {\bibinfo {volume} {80}},\ \bibinfo {pages} {075104} (\bibinfo {year} {2009})}\BibitemShut {NoStop}%
\bibitem [{\citenamefont {Kusunose}(2006)}]{Kusunose2006}%
  \BibitemOpen
  \bibfield  {author} {\bibinfo {author} {\bibfnamefont {H.}~\bibnamefont {Kusunose}},\ }\href {\doibase 10.1143/JPSJ.75.054713} {\bibfield  {journal} {\bibinfo  {journal} {Journal of the Physical Society of Japan}\ }\textbf {\bibinfo {volume} {75}},\ \bibinfo {pages} {054713} (\bibinfo {year} {2006})},\ \Eprint {http://arxiv.org/abs/https://doi.org/10.1143/JPSJ.75.054713} {https://doi.org/10.1143/JPSJ.75.054713} \BibitemShut {NoStop}%
\bibitem [{\citenamefont {Rohringer}\ \emph {et~al.}(2018)\citenamefont {Rohringer}, \citenamefont {Hafermann}, \citenamefont {Toschi}, \citenamefont {Katanin}, \citenamefont {Antipov}, \citenamefont {Katsnelson}, \citenamefont {Lichtenstein}, \citenamefont {Rubtsov},\ and\ \citenamefont {Held}}]{Rohringer2018}%
  \BibitemOpen
  \bibfield  {author} {\bibinfo {author} {\bibfnamefont {G.}~\bibnamefont {Rohringer}}, \bibinfo {author} {\bibfnamefont {H.}~\bibnamefont {Hafermann}}, \bibinfo {author} {\bibfnamefont {A.}~\bibnamefont {Toschi}}, \bibinfo {author} {\bibfnamefont {A.~A.}\ \bibnamefont {Katanin}}, \bibinfo {author} {\bibfnamefont {A.~E.}\ \bibnamefont {Antipov}}, \bibinfo {author} {\bibfnamefont {M.~I.}\ \bibnamefont {Katsnelson}}, \bibinfo {author} {\bibfnamefont {A.~I.}\ \bibnamefont {Lichtenstein}}, \bibinfo {author} {\bibfnamefont {A.~N.}\ \bibnamefont {Rubtsov}}, \ and\ \bibinfo {author} {\bibfnamefont {K.}~\bibnamefont {Held}},\ }\href {\doibase 10.1103/RevModPhys.90.025003} {\bibfield  {journal} {\bibinfo  {journal} {Rev. Mod. Phys.}\ }\textbf {\bibinfo {volume} {90}},\ \bibinfo {pages} {025003} (\bibinfo {year} {2018})}\BibitemShut {NoStop}%
\bibitem [{DGA()}]{DGAdetails}%
  \BibitemOpen
  \href@noop {} {}\bibinfo {note} {We employ the exact diagonalization method to solve mapped quantum impurity models in DMFT loops, and use 120$\times$120 momentum grids, and following previous studies \cite{Kitatani2019,Kitatani2022}, employ 60(1024) Matsubara points for the positive side of the inner(outer) box, respectively.}\BibitemShut {Stop}%
\bibitem [{\citenamefont {Kitatani}\ \emph {et~al.}(2019)\citenamefont {Kitatani}, \citenamefont {Sch\"afer}, \citenamefont {Aoki},\ and\ \citenamefont {Held}}]{Kitatani2019}%
  \BibitemOpen
  \bibfield  {author} {\bibinfo {author} {\bibfnamefont {M.}~\bibnamefont {Kitatani}}, \bibinfo {author} {\bibfnamefont {T.}~\bibnamefont {Sch\"afer}}, \bibinfo {author} {\bibfnamefont {H.}~\bibnamefont {Aoki}}, \ and\ \bibinfo {author} {\bibfnamefont {K.}~\bibnamefont {Held}},\ }\href {\doibase 10.1103/PhysRevB.99.041115} {\bibfield  {journal} {\bibinfo  {journal} {Phys. Rev. B}\ }\textbf {\bibinfo {volume} {99}},\ \bibinfo {pages} {041115} (\bibinfo {year} {2019})}\BibitemShut {NoStop}%
\bibitem [{\citenamefont {{Kitatani}}\ \emph {et~al.}(2020)\citenamefont {{Kitatani}}, \citenamefont {{Si}}, \citenamefont {{Janson}}, \citenamefont {{Arita}}, \citenamefont {{Zhong}},\ and\ \citenamefont {{Held}}}]{Kitatani2020}%
  \BibitemOpen
  \bibfield  {author} {\bibinfo {author} {\bibfnamefont {M.}~\bibnamefont {{Kitatani}}}, \bibinfo {author} {\bibfnamefont {L.}~\bibnamefont {{Si}}}, \bibinfo {author} {\bibfnamefont {O.}~\bibnamefont {{Janson}}}, \bibinfo {author} {\bibfnamefont {R.}~\bibnamefont {{Arita}}}, \bibinfo {author} {\bibfnamefont {Z.}~\bibnamefont {{Zhong}}}, \ and\ \bibinfo {author} {\bibfnamefont {K.}~\bibnamefont {{Held}}},\ }\href {\doibase 10.1038/s41535-020-00260-y} {\bibfield  {journal} {\bibinfo  {journal} {npj Quantum Materials}\ }\textbf {\bibinfo {volume} {5}},\ \bibinfo {pages} {59} (\bibinfo {year} {2020})}\BibitemShut {NoStop}%
\bibitem [{\citenamefont {Kitatani}\ \emph {et~al.}(2023)\citenamefont {Kitatani}, \citenamefont {Si}, \citenamefont {Worm}, \citenamefont {Tomczak}, \citenamefont {Arita},\ and\ \citenamefont {Held}}]{Kitatani2023}%
  \BibitemOpen
  \bibfield  {author} {\bibinfo {author} {\bibfnamefont {M.}~\bibnamefont {Kitatani}}, \bibinfo {author} {\bibfnamefont {L.}~\bibnamefont {Si}}, \bibinfo {author} {\bibfnamefont {P.}~\bibnamefont {Worm}}, \bibinfo {author} {\bibfnamefont {J.~M.}\ \bibnamefont {Tomczak}}, \bibinfo {author} {\bibfnamefont {R.}~\bibnamefont {Arita}}, \ and\ \bibinfo {author} {\bibfnamefont {K.}~\bibnamefont {Held}},\ }\href {\doibase 10.1103/PhysRevLett.130.166002} {\bibfield  {journal} {\bibinfo  {journal} {Phys. Rev. Lett.}\ }\textbf {\bibinfo {volume} {130}},\ \bibinfo {pages} {166002} (\bibinfo {year} {2023})}\BibitemShut {NoStop}%
\bibitem [{\citenamefont {Rohringer}\ \emph {et~al.}(2011)\citenamefont {Rohringer}, \citenamefont {Toschi}, \citenamefont {Katanin},\ and\ \citenamefont {Held}}]{Rohringer2011}%
  \BibitemOpen
  \bibfield  {author} {\bibinfo {author} {\bibfnamefont {G.}~\bibnamefont {Rohringer}}, \bibinfo {author} {\bibfnamefont {A.}~\bibnamefont {Toschi}}, \bibinfo {author} {\bibfnamefont {A.}~\bibnamefont {Katanin}}, \ and\ \bibinfo {author} {\bibfnamefont {K.}~\bibnamefont {Held}},\ }\href {\doibase 10.1103/PhysRevLett.107.256402} {\bibfield  {journal} {\bibinfo  {journal} {Phys. Rev. Lett.}\ }\textbf {\bibinfo {volume} {107}},\ \bibinfo {pages} {256402} (\bibinfo {year} {2011})}\BibitemShut {NoStop}%
\bibitem [{\citenamefont {Sch\"afer}\ \emph {et~al.}(2017)\citenamefont {Sch\"afer}, \citenamefont {Katanin}, \citenamefont {Held},\ and\ \citenamefont {Toschi}}]{Schaefer2017}%
  \BibitemOpen
  \bibfield  {author} {\bibinfo {author} {\bibfnamefont {T.}~\bibnamefont {Sch\"afer}}, \bibinfo {author} {\bibfnamefont {A.~A.}\ \bibnamefont {Katanin}}, \bibinfo {author} {\bibfnamefont {K.}~\bibnamefont {Held}}, \ and\ \bibinfo {author} {\bibfnamefont {A.}~\bibnamefont {Toschi}},\ }\href {\doibase 10.1103/PhysRevLett.119.046402} {\bibfield  {journal} {\bibinfo  {journal} {Phys. Rev. Lett.}\ }\textbf {\bibinfo {volume} {119}},\ \bibinfo {pages} {046402} (\bibinfo {year} {2017})}\BibitemShut {NoStop}%
\bibitem [{\citenamefont {Sch\"afer}\ \emph {et~al.}(2019)\citenamefont {Sch\"afer}, \citenamefont {Katanin}, \citenamefont {Kitatani}, \citenamefont {Toschi},\ and\ \citenamefont {Held}}]{Schaefer2019}%
  \BibitemOpen
  \bibfield  {author} {\bibinfo {author} {\bibfnamefont {T.}~\bibnamefont {Sch\"afer}}, \bibinfo {author} {\bibfnamefont {A.~A.}\ \bibnamefont {Katanin}}, \bibinfo {author} {\bibfnamefont {M.}~\bibnamefont {Kitatani}}, \bibinfo {author} {\bibfnamefont {A.}~\bibnamefont {Toschi}}, \ and\ \bibinfo {author} {\bibfnamefont {K.}~\bibnamefont {Held}},\ }\href {\doibase 10.1103/PhysRevLett.122.227201} {\bibfield  {journal} {\bibinfo  {journal} {Phys. Rev. Lett.}\ }\textbf {\bibinfo {volume} {122}},\ \bibinfo {pages} {227201} (\bibinfo {year} {2019})}\BibitemShut {NoStop}%
\bibitem [{\citenamefont {Kitatani}\ \emph {et~al.}(2025)\citenamefont {Kitatani}, \citenamefont {Sch{\"a}fer}, \citenamefont {Katanin}, \citenamefont {Toschi},\ and\ \citenamefont {Held}}]{Kitatani2025}%
  \BibitemOpen
  \bibfield  {author} {\bibinfo {author} {\bibfnamefont {M.}~\bibnamefont {Kitatani}}, \bibinfo {author} {\bibfnamefont {T.}~\bibnamefont {Sch{\"a}fer}}, \bibinfo {author} {\bibfnamefont {A.}~\bibnamefont {Katanin}}, \bibinfo {author} {\bibfnamefont {A.}~\bibnamefont {Toschi}}, \ and\ \bibinfo {author} {\bibfnamefont {K.}~\bibnamefont {Held}},\ }\href@noop {} {\bibfield  {journal} {\bibinfo  {journal} {arXiv preprint arXiv:2503.09529}\ } (\bibinfo {year} {2025})}\BibitemShut {NoStop}%
\bibitem [{\citenamefont {Pickem}\ \emph {et~al.}(2022)\citenamefont {Pickem}, \citenamefont {Tomczak},\ and\ \citenamefont {Held}}]{Pickem2022}%
  \BibitemOpen
  \bibfield  {author} {\bibinfo {author} {\bibfnamefont {M.}~\bibnamefont {Pickem}}, \bibinfo {author} {\bibfnamefont {J.~M.}\ \bibnamefont {Tomczak}}, \ and\ \bibinfo {author} {\bibfnamefont {K.}~\bibnamefont {Held}},\ }\href {\doibase 10.1103/PhysRevResearch.4.033253} {\bibfield  {journal} {\bibinfo  {journal} {Phys. Rev. Res.}\ }\textbf {\bibinfo {volume} {4}},\ \bibinfo {pages} {033253} (\bibinfo {year} {2022})}\BibitemShut {NoStop}%
\bibitem [{\citenamefont {Bickers}\ \emph {et~al.}(1989)\citenamefont {Bickers}, \citenamefont {Scalapino},\ and\ \citenamefont {White}}]{Bickers1989}%
  \BibitemOpen
  \bibfield  {author} {\bibinfo {author} {\bibfnamefont {N.~E.}\ \bibnamefont {Bickers}}, \bibinfo {author} {\bibfnamefont {D.~J.}\ \bibnamefont {Scalapino}}, \ and\ \bibinfo {author} {\bibfnamefont {S.~R.}\ \bibnamefont {White}},\ }\href {\doibase 10.1103/PhysRevLett.62.961} {\bibfield  {journal} {\bibinfo  {journal} {Phys. Rev. Lett.}\ }\textbf {\bibinfo {volume} {62}},\ \bibinfo {pages} {961} (\bibinfo {year} {1989})}\BibitemShut {NoStop}%
\bibitem [{SM()}]{SM}%
  \BibitemOpen
  \href@noop {} {}\bibinfo {note} {Supplemental information are available at XXX.}\BibitemShut {Stop}%
\bibitem [{\citenamefont {Huang}\ \emph {et~al.}(2019)\citenamefont {Huang}, \citenamefont {Vaezi}, \citenamefont {Nussinov},\ and\ \citenamefont {Vaezi}}]{Huang2019}%
  \BibitemOpen
  \bibfield  {author} {\bibinfo {author} {\bibfnamefont {E.~W.}\ \bibnamefont {Huang}}, \bibinfo {author} {\bibfnamefont {M.-S.}\ \bibnamefont {Vaezi}}, \bibinfo {author} {\bibfnamefont {Z.}~\bibnamefont {Nussinov}}, \ and\ \bibinfo {author} {\bibfnamefont {A.}~\bibnamefont {Vaezi}},\ }\href {\doibase 10.1103/PhysRevB.99.235128} {\bibfield  {journal} {\bibinfo  {journal} {Phys. Rev. B}\ }\textbf {\bibinfo {volume} {99}},\ \bibinfo {pages} {235128} (\bibinfo {year} {2019})}\BibitemShut {NoStop}%
\bibitem [{\citenamefont {Dzyaloshinskii}(2003)}]{Dzyaloshinskii2003}%
  \BibitemOpen
  \bibfield  {author} {\bibinfo {author} {\bibfnamefont {I.}~\bibnamefont {Dzyaloshinskii}},\ }\href {\doibase 10.1103/PhysRevB.68.085113} {\bibfield  {journal} {\bibinfo  {journal} {Phys. Rev. B}\ }\textbf {\bibinfo {volume} {68}},\ \bibinfo {pages} {085113} (\bibinfo {year} {2003})}\BibitemShut {NoStop}%
\bibitem [{\citenamefont {Stanescu}\ and\ \citenamefont {Kotliar}(2006)}]{Stanescu2006}%
  \BibitemOpen
  \bibfield  {author} {\bibinfo {author} {\bibfnamefont {T.~D.}\ \bibnamefont {Stanescu}}\ and\ \bibinfo {author} {\bibfnamefont {G.}~\bibnamefont {Kotliar}},\ }\href {\doibase 10.1103/PhysRevB.74.125110} {\bibfield  {journal} {\bibinfo  {journal} {Phys. Rev. B}\ }\textbf {\bibinfo {volume} {74}},\ \bibinfo {pages} {125110} (\bibinfo {year} {2006})}\BibitemShut {NoStop}%
\bibitem [{\citenamefont {Stanescu}\ \emph {et~al.}(2007)\citenamefont {Stanescu}, \citenamefont {Phillips},\ and\ \citenamefont {Choy}}]{Stanescu2007}%
  \BibitemOpen
  \bibfield  {author} {\bibinfo {author} {\bibfnamefont {T.~D.}\ \bibnamefont {Stanescu}}, \bibinfo {author} {\bibfnamefont {P.}~\bibnamefont {Phillips}}, \ and\ \bibinfo {author} {\bibfnamefont {T.-P.}\ \bibnamefont {Choy}},\ }\href {\doibase 10.1103/PhysRevB.75.104503} {\bibfield  {journal} {\bibinfo  {journal} {Phys. Rev. B}\ }\textbf {\bibinfo {volume} {75}},\ \bibinfo {pages} {104503} (\bibinfo {year} {2007})}\BibitemShut {NoStop}%
\bibitem [{\citenamefont {Sakai}\ \emph {et~al.}(2009)\citenamefont {Sakai}, \citenamefont {Motome},\ and\ \citenamefont {Imada}}]{Sakai2009}%
  \BibitemOpen
  \bibfield  {author} {\bibinfo {author} {\bibfnamefont {S.}~\bibnamefont {Sakai}}, \bibinfo {author} {\bibfnamefont {Y.}~\bibnamefont {Motome}}, \ and\ \bibinfo {author} {\bibfnamefont {M.}~\bibnamefont {Imada}},\ }\href {\doibase 10.1103/PhysRevLett.102.056404} {\bibfield  {journal} {\bibinfo  {journal} {Phys. Rev. Lett.}\ }\textbf {\bibinfo {volume} {102}},\ \bibinfo {pages} {056404} (\bibinfo {year} {2009})}\BibitemShut {NoStop}%
\bibitem [{\citenamefont {Sakai}(2023)}]{Sakai2023}%
  \BibitemOpen
  \bibfield  {author} {\bibinfo {author} {\bibfnamefont {S.}~\bibnamefont {Sakai}},\ }\href {\doibase 10.7566/JPSJ.92.092001} {\bibfield  {journal} {\bibinfo  {journal} {Journal of the Physical Society of Japan}\ }\textbf {\bibinfo {volume} {92}},\ \bibinfo {pages} {092001} (\bibinfo {year} {2023})},\ \Eprint {http://arxiv.org/abs/https://doi.org/10.7566/JPSJ.92.092001} {https://doi.org/10.7566/JPSJ.92.092001} \BibitemShut {NoStop}%
\bibitem [{\citenamefont {Kaufmann}\ and\ \citenamefont {Held}(2023)}]{Kaufmann2023}%
  \BibitemOpen
  \bibfield  {author} {\bibinfo {author} {\bibfnamefont {J.}~\bibnamefont {Kaufmann}}\ and\ \bibinfo {author} {\bibfnamefont {K.}~\bibnamefont {Held}},\ }\href {\doibase https://doi.org/10.1016/j.cpc.2022.108519} {\bibfield  {journal} {\bibinfo  {journal} {Computer Physics Communications}\ }\textbf {\bibinfo {volume} {282}},\ \bibinfo {pages} {108519} (\bibinfo {year} {2023})}\BibitemShut {NoStop}%
\bibitem [{ana()}]{anacont}%
  \BibitemOpen
  \href@noop {} {}\bibinfo {note} {As in previous studies \cite{Worm2024}, a minimum imaginary part of the self-energy, $\gamma=-0.02t$, was imposed (simply added in this study) to obtain spectral weight, which does not affect the overall shape.}\BibitemShut {Stop}%
\bibitem [{\citenamefont {Hatsugai}\ and\ \citenamefont {Kohmoto}(1992)}]{Hatsugai1992}%
  \BibitemOpen
  \bibfield  {author} {\bibinfo {author} {\bibfnamefont {Y.}~\bibnamefont {Hatsugai}}\ and\ \bibinfo {author} {\bibfnamefont {M.}~\bibnamefont {Kohmoto}},\ }\href@noop {} {\bibfield  {journal} {\bibinfo  {journal} {Journal of the Physical Society of Japan}\ }\textbf {\bibinfo {volume} {61}},\ \bibinfo {pages} {2056} (\bibinfo {year} {1992})}\BibitemShut {NoStop}%
\bibitem [{\citenamefont {Kitatani}\ \emph {et~al.}(2022)\citenamefont {Kitatani}, \citenamefont {Arita}, \citenamefont {Schäfer},\ and\ \citenamefont {Held}}]{Kitatani2022}%
  \BibitemOpen
  \bibfield  {author} {\bibinfo {author} {\bibfnamefont {M.}~\bibnamefont {Kitatani}}, \bibinfo {author} {\bibfnamefont {R.}~\bibnamefont {Arita}}, \bibinfo {author} {\bibfnamefont {T.}~\bibnamefont {Schäfer}}, \ and\ \bibinfo {author} {\bibfnamefont {K.}~\bibnamefont {Held}},\ }\href {\doibase 10.1088/2515-7639/ac7e6d} {\bibfield  {journal} {\bibinfo  {journal} {Journal of Physics: Materials}\ }\textbf {\bibinfo {volume} {5}},\ \bibinfo {pages} {034005} (\bibinfo {year} {2022})}\BibitemShut {NoStop}%
\bibitem [{\citenamefont {Worm}\ \emph {et~al.}(2024)\citenamefont {Worm}, \citenamefont {Reitner}, \citenamefont {Held},\ and\ \citenamefont {Toschi}}]{Worm2024}%
  \BibitemOpen
  \bibfield  {author} {\bibinfo {author} {\bibfnamefont {P.}~\bibnamefont {Worm}}, \bibinfo {author} {\bibfnamefont {M.}~\bibnamefont {Reitner}}, \bibinfo {author} {\bibfnamefont {K.}~\bibnamefont {Held}}, \ and\ \bibinfo {author} {\bibfnamefont {A.}~\bibnamefont {Toschi}},\ }\href {\doibase 10.1103/PhysRevLett.133.166501} {\bibfield  {journal} {\bibinfo  {journal} {Phys. Rev. Lett.}\ }\textbf {\bibinfo {volume} {133}},\ \bibinfo {pages} {166501} (\bibinfo {year} {2024})}\BibitemShut {NoStop}%
\end{thebibliography}%


\begin{thebibliography}{3}
\expandafter\ifx\csname natexlab\endcsname\relax\def\natexlab#1{#1}\fi
\expandafter\ifx\csname bibnamefont\endcsname\relax
  \def\bibnamefont#1{#1}\fi
\expandafter\ifx\csname bibfnamefont\endcsname\relax
  \def\bibfnamefont#1{#1}\fi
\expandafter\ifx\csname citenamefont\endcsname\relax
  \def\citenamefont#1{#1}\fi
\expandafter\ifx\csname url\endcsname\relax
  \def\url#1{\texttt{#1}}\fi
\expandafter\ifx\csname urlprefix\endcsname\relax\def\urlprefix{URL }\fi
\providecommand{\bibinfo}[2]{#2}
\providecommand{\eprint}[2][]{\url{#2}}

\bibitem[{pol()}]{pole}
\bibinfo{note}{While the integration can be done without this assumption by reformulating $\frac{1}{A}\int d\theta/(1+\sin\theta B/A)$, We found the difference is minor numerically, and we here use the Eq.~(S.8) for simplicity.}

\bibitem[{\citenamefont{Schmalian et~al.}(1999)\citenamefont{Schmalian, Pines, and Stojkovi\ifmmode~\acute{c}\else \'{c}\fi{}}}]{Schmalian1999}
\bibinfo{author}{\bibfnamefont{J.}~\bibnamefont{Schmalian}}, \bibinfo{author}{\bibfnamefont{D.}~\bibnamefont{Pines}}, \bibnamefont{and} \bibinfo{author}{\bibfnamefont{B.}~\bibnamefont{Stojkovi\ifmmode~\acute{c}\else \'{c}\fi{}}}, \bibinfo{journal}{Phys. Rev. B} \textbf{\bibinfo{volume}{60}}, \bibinfo{pages}{667} (\bibinfo{year}{1999}), \urlprefix\url{https://link.aps.org/doi/10.1103/PhysRevB.60.667}.

\bibitem[{\citenamefont{Katanin}(2005)}]{Katanin2005b}
\bibinfo{author}{\bibfnamefont{A.~A.} \bibnamefont{Katanin}}, \bibinfo{journal}{Phys. Rev. B} \textbf{\bibinfo{volume}{72}}, \bibinfo{pages}{035111} (\bibinfo{year}{2005}), \urlprefix\url{https://link.aps.org/doi/10.1103/PhysRevB.72.035111}.

\end{thebibliography}
\end{document}


\title{Supplementary material for ``Luttinger surface and exchange splitting induced by ferromagnetic fluctuations"
}

\author{Motoharu Kitatani}
\affiliation{Department of Material Science, University of Hyogo, Ako, Hyogo 678-1297, Japan}

\author{Yusuke Nomura}
\affiliation{Institute for Materials Research (IMR), Tohoku University, Katahira, Aoba-ku, Sendai 980-8577, Japan}
\affiliation{Advanced Institute for Materials Research (WPI-AIMR), Tohoku University, Katahira, Aoba-ku, Sendai 980-8577, Japan}

\author{Shiro Sakai}
\affiliation{Physics Division, Sophia University, Chiyoda-ku, Tokyo 102-8554, Japan}
\affiliation{RIKEN Center for Emergent Matter Sciences (CEMS), Wako, Saitama, 351-0198, Japan}

\author{Ryotaro Arita}
\affiliation{RIKEN Center for Emergent Matter Sciences (CEMS), Wako, Saitama, 351-0198, Japan}
\affiliation{Department of Physics, The University of Tokyo, Hongo, Tokyo, 113-8656, Japan}

\date{\today}

\maketitle

\section{Additional results of Fermi surfaces}
\begin{figure}[htbp]
        \centering
        \includegraphics[width=0.75\linewidth,angle=0]{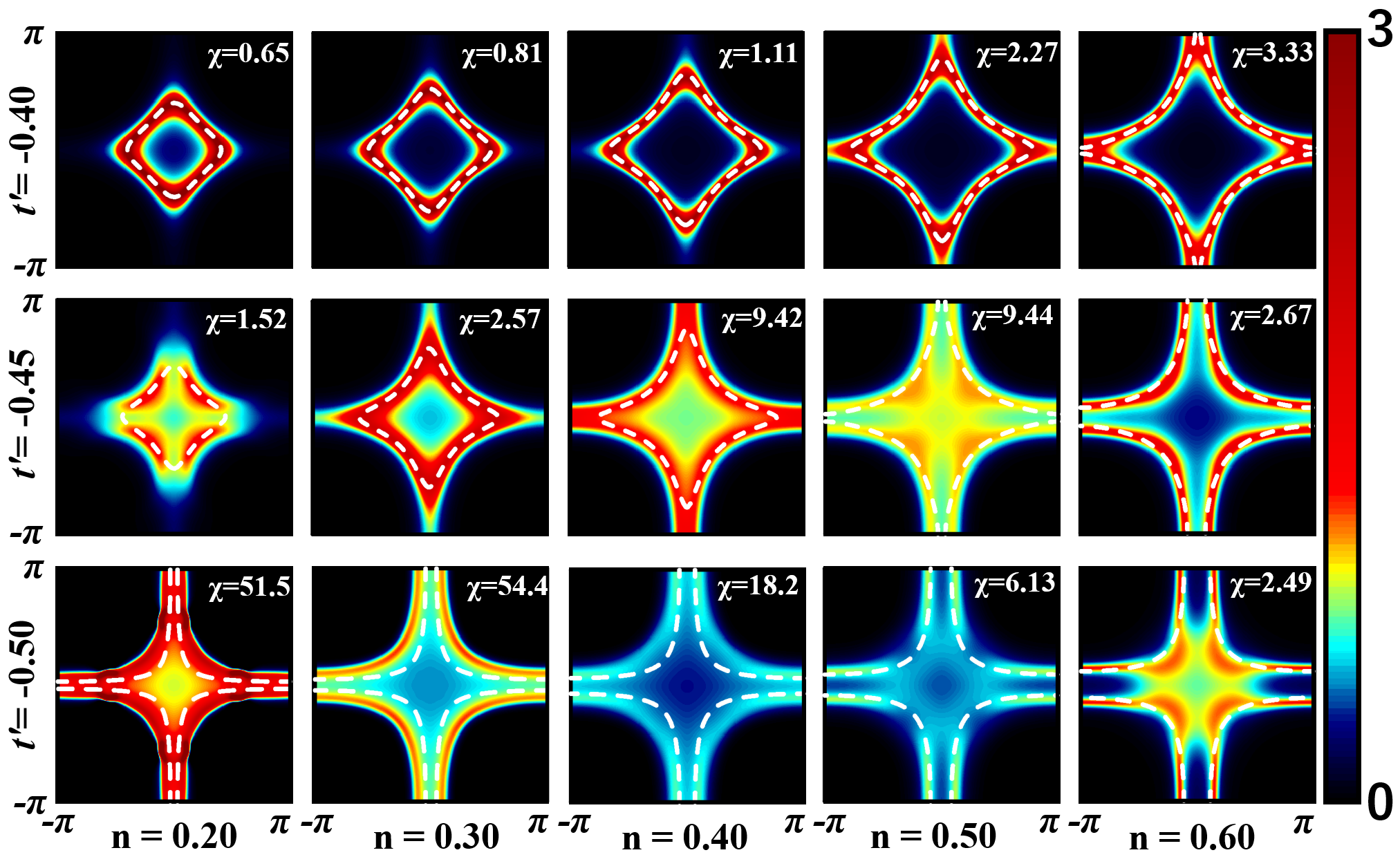}
        \caption{Momentum dependence of the imaginary part of the Green's function $-\Im G/\pi$ at the lowest Matsubara frequency in D$\Gamma$A for $t^{\prime}= -0.40t,\ -0.45t,\ -0.50t$, $n=0.20,\ 0.30,\ 0.40,\ 0.50,\ 0.60$, $T=0.02t$ and $U=4t$. The value of the ferromagnetic susceptibility $\chi_{\rm s}(\mathbf{k}=0,\omega=0)$ is also shown in each figure. Dashed curves indicate the non-interacting Fermi surface for each system.}
        \label{fig:U4}
\end{figure}

\begin{figure}[htbp]
        \centering
        \includegraphics[width=0.75\linewidth,angle=0]{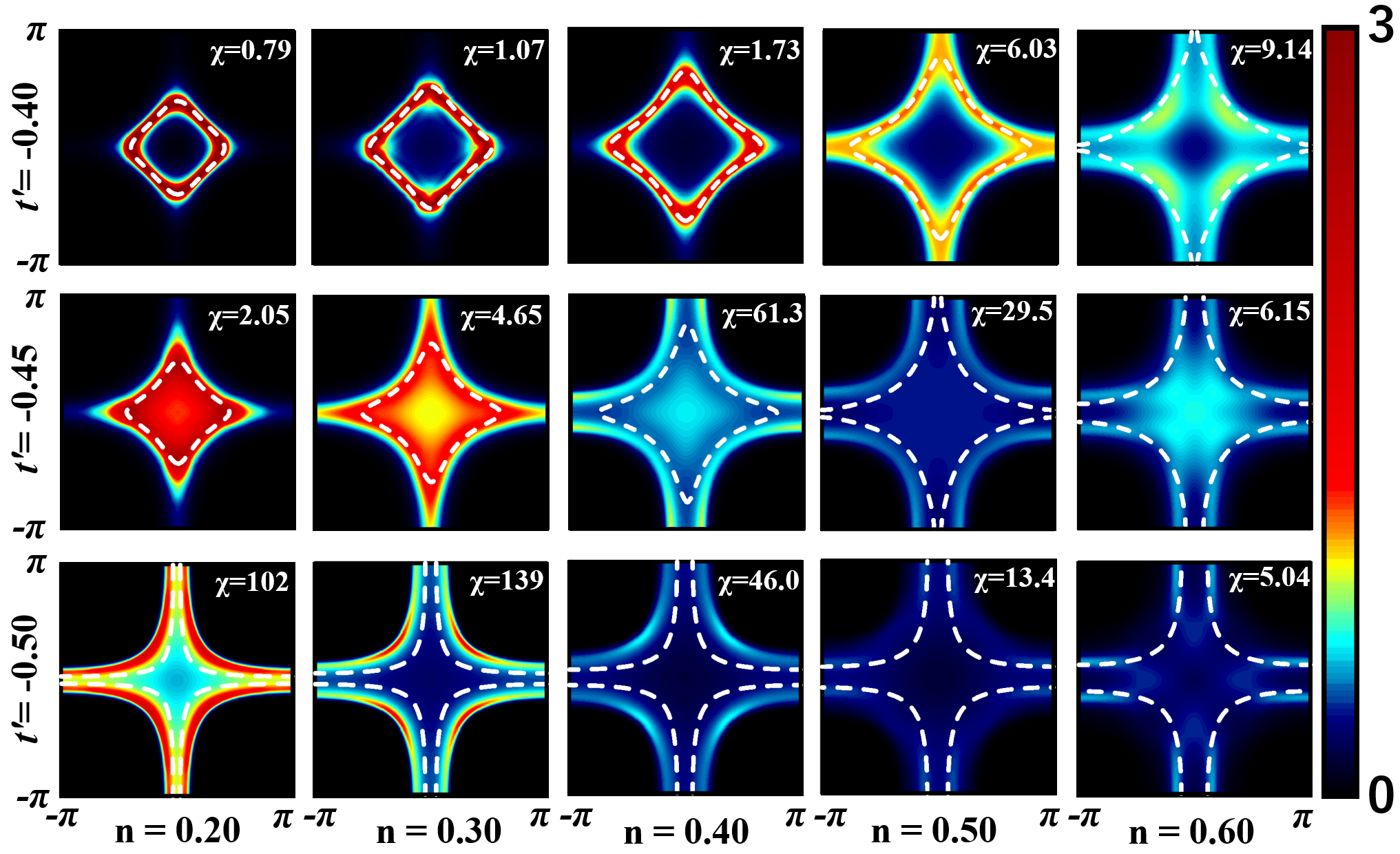}
        \caption{Same plot as Fig.~\ref{fig:U4}, but for $U=6t$.}
        \label{fig:U6}
\end{figure}

\begin{figure}[htbp]
        \centering
        \includegraphics[width=0.75\linewidth,angle=0]{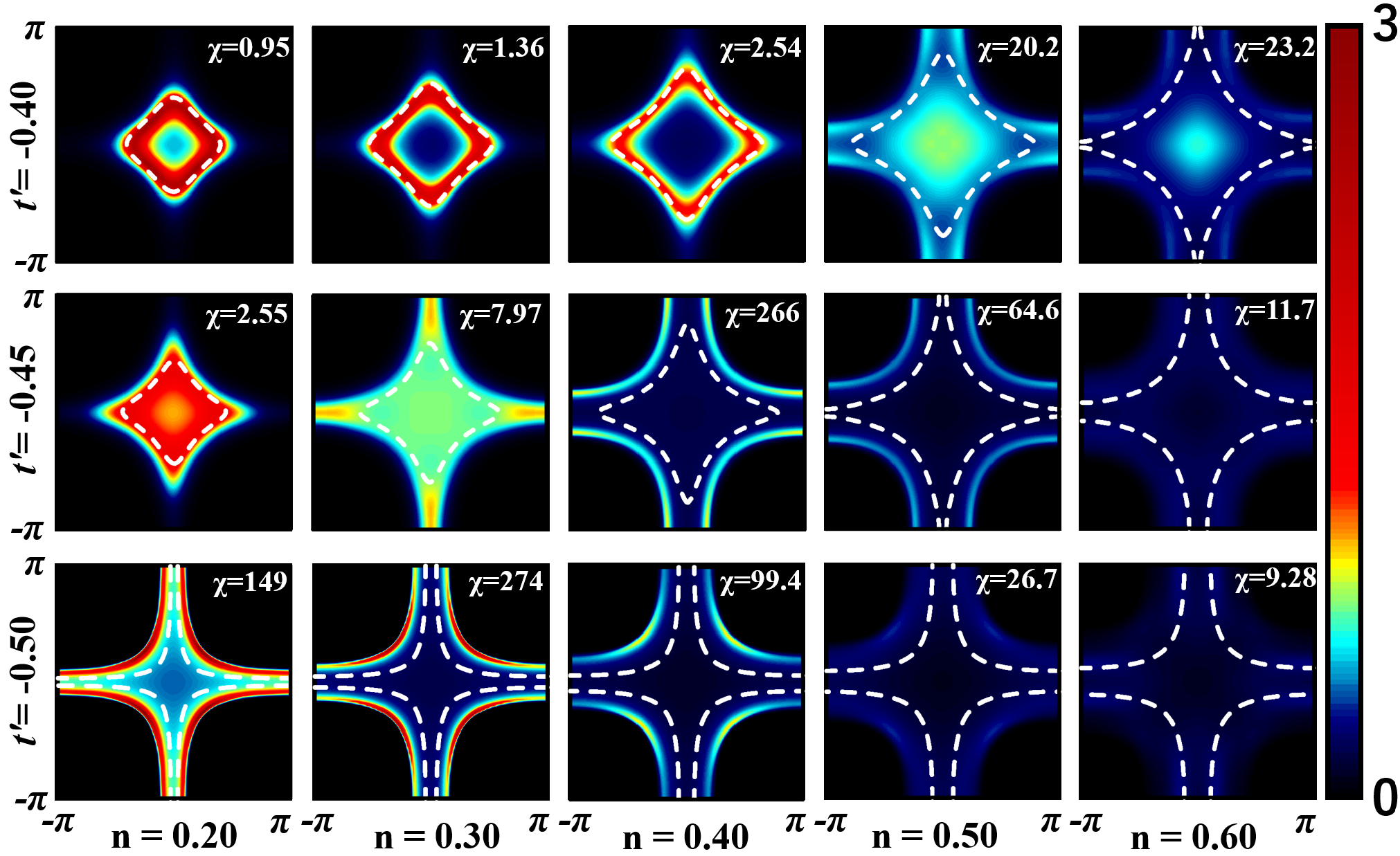}
        \caption{Same plot as Fig.~\ref{fig:U4}, but for $U=8t$.}
        \label{fig:U8}
\end{figure}

Here, we show the D$\Gamma$A Fermi surfaces, similar to Fig.~2 of the main text, but for a wider range of parameters: $U=4t$ (Fig.~\ref{fig:U4}), $U=6t$ (Fig.~\ref{fig:U6}) and $U=8t$ (Fig.~\ref{fig:U8}). We find that the discussion in the main text also applies here. Specifically, the Fermi surface follows the non-interacting (white dashed) line with weak ferromagnetic fluctuations, and the Fermi surface expands as the ferromagnetic fluctuation become stronger. 

\begin{figure}[htbp]
        \centering
        \includegraphics[width=0.6\linewidth,angle=0]{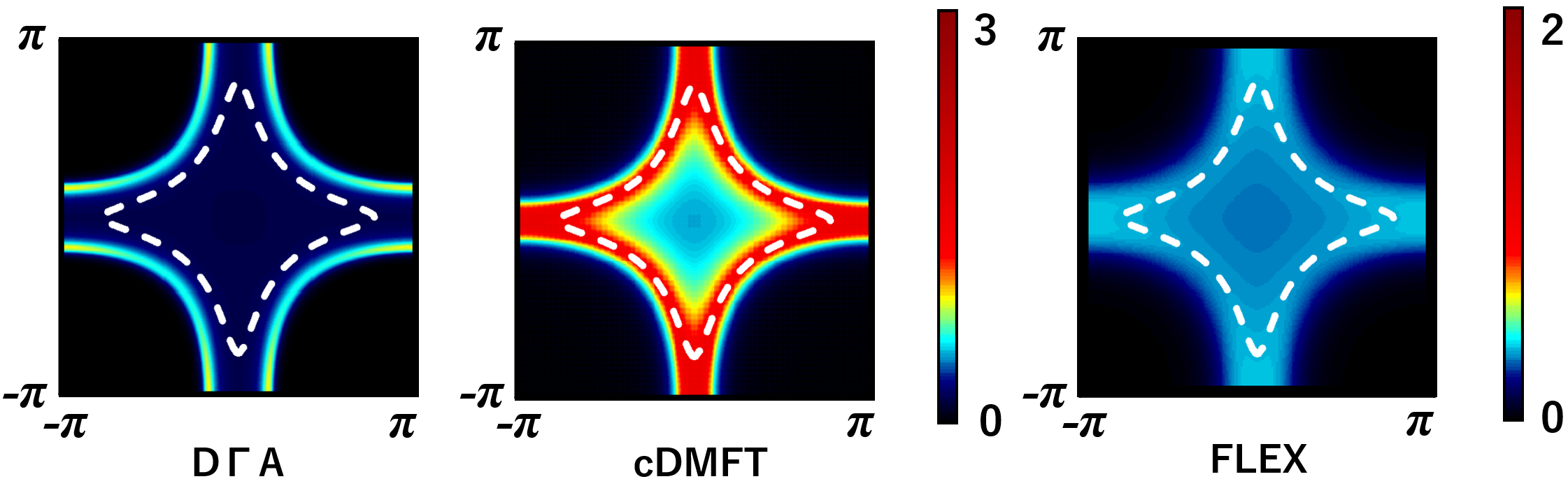}
        \caption{Comparison among different methods for $U=8t, t'=-0.45t$ and $n=0.40$.}
        \label{fig:comparison}
\end{figure}

We also compare the results obtained with different approaches at the fixed parameters ($U=8t,\ t'=-0.45t,\ n=0.40$) in Fig.~\ref{fig:comparison}. Similar to the pseudogap near the half-filling, FLEX shows a broadened spectrum along the non-interacting line. We also observed that the 4-site cDMFT study does not show the change of the Fermi surface, which suggests the long-range fluctuation effect is crucial at this parameter, similar to the antiferromagnetic pseudogap in the half-filled Hubbard model for weak to intermediate coupling regions.


\section{Derivation of the analytical formula}
In this section, we show how we obtained the self-energy formula from the simple assumption $\Sigma=1/\beta\sum_q G(k+q)W(q).$
We expand the momentum dependence of the non-interacting Green's function $G(k+q)$ in powers of $\mathbf{q}$ around $\mathbf{k}$ as
\begin{align}
\Sigma(k) & 
\sim \frac{1}{A\beta} \int \frac{d^2 \mathbf{q}}{(2 \pi)^2} \frac{1}{G_0(k)^{-1} - (\epsilon_{\mathbf{k+q}} - \epsilon_{\mathbf{k}})} \frac{1}{|\mathbf{q}|^2 + \xi^{-2}}, \\
\approx & \frac{1}{A\beta} \int \frac{d^2 \mathbf{q}}{(2 \pi)^2} \frac{1}{G_0(k)^{-1} - \left( 
\frac{\partial \epsilon}{\partial k_x} q_x
+\frac{\partial \epsilon}{\partial k_y} q_y
+\frac{1}{2} \frac{\partial^2 \epsilon}{\partial k_x^2} q_x^2 
+\frac{\partial^2 \epsilon}{\partial k_x \partial k_y} q_x q_y 
+\frac{1}{2} \frac{\partial^2 \epsilon}{\partial k_y^2} q_y^2
\right)} \frac{1}{|\mathbf{q}|^2 + \xi^{-2}}, \label{eq:int-Gchi} \\
\approx& \frac{1}{A\beta}
\int_0^{\delta} \frac{dq}{(2\pi)^2} \int_0^{2\pi} d\theta \frac{q}{q^2 + \xi^{-2}} \\ \nonumber
&\times
\frac{1}{G_0(k)^{-1} - \left[ 
q\left(
\frac{\partial \epsilon}{\partial k_x} \cos\theta + \frac{\partial \epsilon}{\partial k_y} \sin\theta
\right)
+\frac{q^2}{4}\left(
\frac{\partial^2 \epsilon}{\partial k_x^2}
+\frac{\partial^2 \epsilon}{\partial k_y^2}
\right)
+\frac{q^2}{4}\left(
\frac{\partial^2 \epsilon}{\partial k_x^2}
-\frac{\partial^2 \epsilon}{\partial k_y^2}
\right) \cos 2\theta
+\frac{q^2}{2}
\frac{\partial^2 \epsilon}{\partial k_x \partial k_y} \sin 2\theta
\right]}.
\end{align}


Taking the lowest order of each of the constant part and the $\cos\theta,\sin\theta$ part, we obtain
\begin{align}
\Sigma(k)
&\sim\frac{1}{A\beta}
\int_0^{\delta} \frac{dq}{(2\pi)^2} \int_0^{2\pi} d\theta \frac{q}{q^2 + \xi^{-2}}
\frac{1}{G_0(k)^{-1}
-\frac{q^2}{4}\left(
\frac{\partial^2 \epsilon}{\partial k_x^2}
+\frac{\partial^2 \epsilon}{\partial k_y^2}
\right)
-q \sqrt{
\left(\frac{\partial \epsilon}{\partial k_x}\right)^2
+\left(\frac{\partial \epsilon}{\partial k_y}\right)^2
}\sin\theta
}, \nonumber\\
&=\frac{1}{A\beta}
\int_0^{\delta} \frac{dq}{(2\pi)^2}
\frac{q G_0(k)}{q^2 + \xi^{-2}}
\int_0^{2\pi} d\theta
\frac{1}{1
-\frac{q^2}{4}\Delta\epsilon G_0(k)
-q |\nabla \epsilon| G_0(k) \sin\theta
}.
\end{align}

For $\theta$ integration,

\begin{align}
\int_0^{2\pi} \frac{d\theta}{A + B \sin\theta} 
&= \oint_{|z|=1} \frac{2dz}{Bz^2 + 2iAz - B} \quad (z \equiv e^{i\theta}), \\
&= \oint_{|z|=1} \frac{2dz}{B(z - \alpha)(z - \beta)} \quad (\alpha, \beta = \frac{-A \pm \sqrt{A^2 - B^2}}{B}i), \\
&= \frac{2\pi}{\sqrt{A^2 - B^2}} \quad 
\left( A \sim 1 \right),
\end{align}
where $A=1-q^2\Delta\epsilon G_0(k)/4,B=q |\nabla \epsilon| G_0(k).$
In the final step, we define the square root as $\Re\sqrt{X}>0$ for any complex number $X$. Assuming the effect of the second order $q^2$ term is weak \cite{pole}, we pick the $z=\alpha$ as the pole within the integration region since $|-1+\sqrt{1-B^2}|<|B|<|1+\sqrt{1-B^2}|$. 

\begin{figure}[htbp]
        \centering
        \includegraphics[width=\linewidth,angle=0]{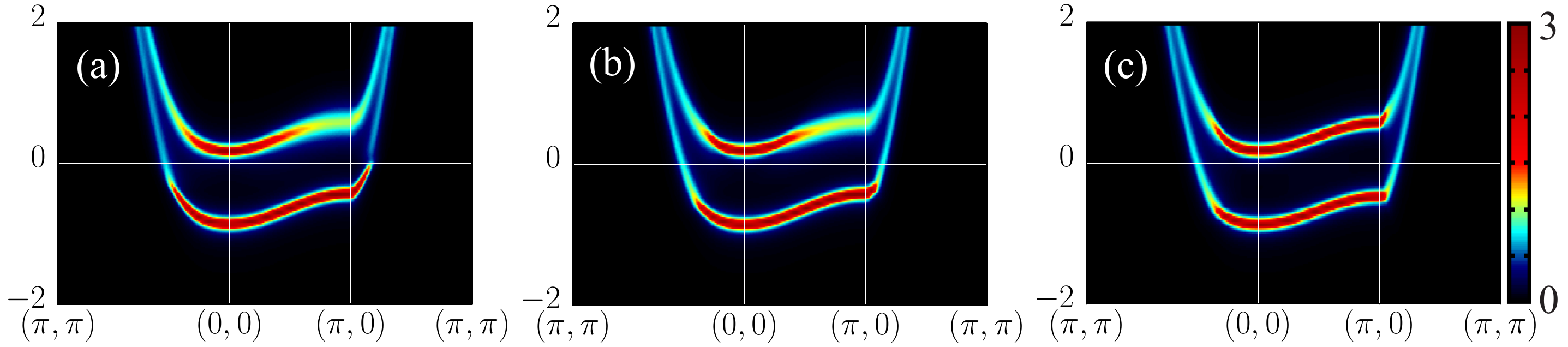}
        \caption{Spectrum for $\xi=10,\ A\beta t^2=1.25,\ t^{\prime}=-0.45t,\mu=-1.850t$ obtained by (a) numerical integration and (b, c) the analytical expression [Eq.~(S.12)]. In (c), the second-order derivative of the dispersion is neglected for comparison.}
        \label{fig:analyticalformula}
\end{figure}

Thus, we obtained the self-energy as
\begin{align}
\Sigma(k) 
&\sim \frac{1}{A\beta} \int_0^\delta \frac{dq}{(2\pi)^2} 
\frac{q G_0(k)}{q^2 + \xi^{-2}}
\frac{2\pi}{\sqrt{ \left(1 - \frac{q^2 \Delta \epsilon}{4} G_0(k) \right)^2 - \left(q |\nabla \epsilon|G_0(k) \right)^2 }}, \\
&= \frac{G_0(k)}{4\pi A\beta} \int_0^\delta dq \frac{2q}{q^2+\xi^{-2}} \frac{1}{\sqrt{1 - \alpha_k q^2 + O(q^4)}} \quad (\alpha_k \equiv |\nabla \epsilon|^2 G_0(k)^2 + \frac{\Delta \epsilon}{2} G_0(k)), \\
&\approx \frac{G_0(k)}{4\pi A\beta} \left[
\frac{2}{\sqrt{1+\alpha_k \xi^{-2}}}\log{
\frac{\sqrt{\xi^{-2}+q^2}}
{\sqrt{1+\alpha_k \xi^{-2}}+\sqrt{1-\alpha_k q^2}}
}
\right]_0^\delta, \\
&=\frac{1}{4\pi^2 A\beta}\frac{2\pi G_0(k)}{\sqrt{1+\alpha_k \xi^{-2}}}
\log{ \left[
\sqrt{\delta^2\xi^2+1}\left(
\frac{\sqrt{1+\alpha_k\xi^{-2}}+1}{\sqrt{1+\alpha_k\xi^{-2}}+\sqrt{1-\alpha_k\delta^2}}
\right)
\right] }.
\label{eq:analyticalformula}
\end{align}
This gives a similar spectrum with Eq.~(14) of Ref.~\cite{Schmalian1999}, but here we include terms up to the second order derivative of the dispersion. In Fig.~(\ref{fig:analyticalformula}), we compare the spectra obtained from the numerical integration of Eq.~\ref{eq:int-Gchi} and from the analytical formula in Eq.~(\ref{eq:analyticalformula}). We find that the second-order correction gives an asymmetric spectral weight above and below the pseudogap. Also, we note that the logarithmic term disappears in the limit of $\xi \rightarrow \infty$ while keeping $\Re\alpha \xi^{-2} \gg 1$ consistent with Ref.~\cite{Katanin2005b}.

\bibliography{main}